\documentclass[aps,pre,twocolumn,groupedaddress]{revtex4-1}

\usepackage{graphicx}
\usepackage{amssymb}
\usepackage{amsmath}

\newcommand \Ubend{U_{\mathrm{bend}}}
\newcommand \Ustrain{U_{\mathrm{strain}}}
\newcommand \Usub{U_{\mathrm{sub}}}
\newcommand \Ubare{U_{\mathrm{bare}}}

\newcommand \q{{{\theta}}}

\newcommand \ru{{{\rm{u}}}}

\newcommand{\eoK}{(\epsilon\chi)}
\newcommand{\eK}{(\epsilon/\chi)}

\newcommand \beq{\begin{equation}}
\newcommand \eeq{\end{equation}}

\newcommand \srr{\sigma_{rr}}
\newcommand \sqq{\sigma_{\theta\theta}}
\newcommand \srq{\sigma_{r\theta}}
\newcommand \srraxi{\sigma_{rr}^{(\mathrm{bare})}}
\newcommand \sqqaxi{\sigma_{\theta\theta}^{(\mathrm{bare})}}

\newcommand \Keff{K_{\mathrm{eff}}}
\newcommand \Ksub{K_{\mathrm{sub}}}

\newcommand \Kcurv{K_{\mathrm{curv}}}
\newcommand \Gsub{G_{\mathrm{sub}}}
\newcommand \Gtar{G_{\mathrm{tar}}}
\newcommand \usub{u_{\mathrm{sub}}}
\newcommand \ubend{u_{\mathrm{bend}}}
\newcommand \uaxi{u^{\mathrm{(bare)}}}
\newcommand \ustrain{u_{\mathrm{strain}}}
\newcommand \varepsilonbare{\varepsilon_{\mathrm{bare}}}
\newcommand \varepsilonres{\varepsilon_{\mathrm{res}}}
\newcommand \uFvK{{u_{\rm FvK}}}
\newcommand \rmu{{\rm u}}
\newcommand \tDelta{{\tilde{\Delta}}}

\newcommand{\xv}{{\bf x}}

\begin{document}

\title{Geometrically incompatible confinement of solids} 

\author{Benny Davidovitch}
\affiliation{Department of Physics, University of Massachusetts, Amherst, MA 01003} 
\author{Yiwei Sun}
\affiliation{Department of Physics, University of Massachusetts, Amherst, MA 01003} 
\author{Gregory M. Grason}
\affiliation{Department of Polymer Science and Engineering, University of Massachusetts, Amherst, MA 01003} 

\date{\today}

\begin{abstract}
The complex morphologies exhibited by spatially confined thin objects have long challenged human efforts to understand and manipulate them, 
from the representation of patterns in draped fabric in Renaissance art to current day efforts to engineer flexible sensors that conform to the human body. We introduce a theoretical principle, broadly generalizing Euler's {\emph{elastica}} -- a core concept of continuum mechanics that invokes the energetic preference of bending over straining a thin solid object and has been widely applied to classical and modern studies of beams and rods. We define a class of {\emph{geometrically incompatible confinement}} problems, whereby the topography imposed on a thin solid body is incompatible with its intrinsic (``target'') metric and, as a consequence of Gauss' {\emph{Theorema Egregium}}, induces strain.  Focusing on a prototypical example of a sheet attached to a spherical substrate, numerical simulations and analytical study demonstrate that the mechanics is governed by a principle, which we call the ``Gauss-Euler {\emph{elastica}}''. This emergent rule states that -- despite the unavoidable strain in such an incompatible  confinement -- the ratio between the energies stored in straining and bending the solid may be arbitrarily small. The Gauss-Euler {\emph{elastica}} underlies a theoretical framework that greatly simplifies the daunting task of solving the highly nonlinear equations that describe thin solids at mechanical equilibrium. This development thus opens new possibilities for attacking a broad class of phenomena governed by the coupling of geometry and mechanics.
\end{abstract}

\maketitle

\section{Introduction}
The spatial confinement of thin, solid objects has been the target of theoretical inquiry since the very early formulations of continuum mechanics. The best known example is ``Galileo's beam'', a problem that is central to solid mechanics, and moreover, to the foundations of variational calculus. Considering the buckling of a wooden beam, Euler recognized that the problem is described by the variational principle \cite{ElasticaReview08}, 
\begin{gather} {\rm Euler's \ {\emph{elastica}}:} \ \ 
\delta \Ubend  = 0;    \  
 {\rm subject\ to} \  \Ustrain = 0 \ ,  
\label{eq:Euler-var} 
\end{gather}
where $\Ubend$ and $\Ustrain$, respectively, are the contributions to the elastic energy of a solid rod due to bending and straining, and $\delta U$ represents the first variation of $U$ with respect to shape.  
Equation~(\ref{eq:Euler-var}), known as ``Euler's {\emph{elastica}}'', is borne out of two principles that intertweave mechanics and geometry. 
First, thin solids are energetically far less costly to bend than to stretch, owing to the fact that the ratio of bending to stretching moduli for a rod (or sheet) of thickness $t$ varies as $\sim t^2$~ \cite{LL86}.  
Second, confining a rod in space -- 
say, by forcing its ends to be closer than its contour length -- can generically be done without straining its centerline.
Thus, by restricting 
configurations to {\it isometric} deformations ({\emph{i.e.}} $\Ustrain = 0$), 
one guarantees that they acquire minimal energy as $t \to 0$, thereby greatly reducing the space of possible equilibria states, making the problem analytically tractable. 

Euler's rule can be extended to two-dimensional (2D) solids under particular circumstances in which the deformed mid-surface can be embedded in 3D isometrically ({\emph{i.e.}} without strain) \cite{Mansfield}, for example, when analyzing the ``developable cones'' realized by gently pushing a xerox paper into a ring \cite{Cerda98}. However, when addressing the generic problem of confinement of a thin solid in 3D space, namely, confinements that require a change in Gaussian curvature of the mid-surface, Gauss' {\emph{Theorema Egregium}} implies that a perfectly isometric deformation of the mid-surface is impossible~\cite{Witten07}, 
revoking the applicability of Euler's variational rule~(\ref{eq:Euler-var}) and requiring, instead, minimization of 
the fully-fledged F\"{o}ppl-von K\'{a}rm\'{a}n (FvK) energy \cite{LL86},
\begin{equation} 
 {\rm FvK:}   \ \ \ \delta[\Ubend + \Ustrain]=0 \ .
\label{eq:FvK-var}
\end{equation}
The unavoidable strain inherent to such {\emph{geometrically incompatible confinement (GIC)}}, underlies both the impossibility of drawing an accurate map of the globe as well as the frustration of gift wrapping a ball.  
A quintessential example of GIC is given by the ``spherical stamping'' experiments of Hure {\emph{et al.}} (Fig.~\ref{fig:1}A), in which a thin sheet (of size $W$ and thickness $t$) is placed in a narrow gap ($2\delta$) between two rigid, concentric spherical shells (of radius $R$) \cite{Hure12}. Here, the positive Gaussian curvature of the confining topography gives rise to complex patterns that cannot be described by Euler's variational rule, necessitating consideration of the highly non-linear interplay between bending and strain in the FvK equations.

\begin{figure*}
\center
\includegraphics[width=1.0\textwidth]{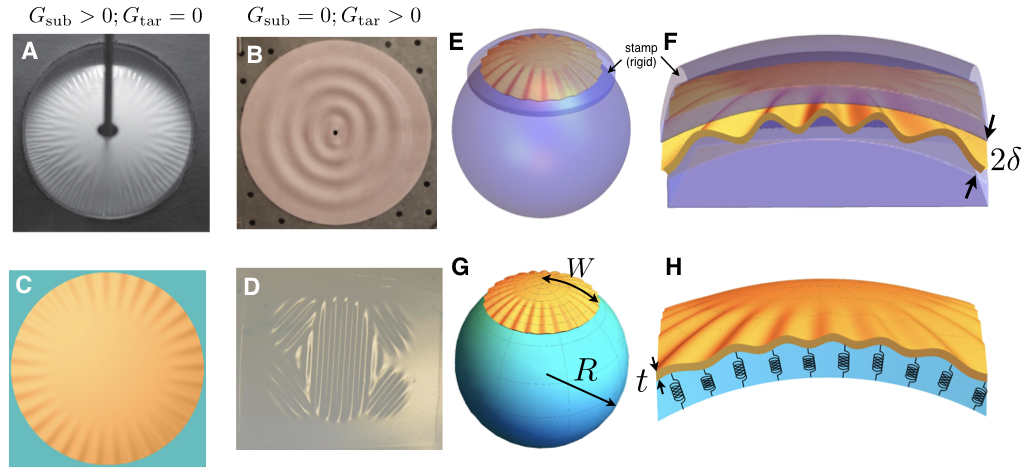}
\caption{\label{fig:1} Examples of geometrically-incompatible confinement (GIC) of thin sheets/shells: (A) circular sheet in a spherical stamp (from ref.~\cite{Hure12}); (B) top view of conical sheet confined between rigid, planar plates (courtesy: E. Sharon); (C) top view of simulated flat sheet confined to a ``spherical Winkler'' substrate
(describe in the text); (D) a polygonal patch cut from a thin spherical shell floated at a (planar) air/liquid interface (from ref.~\cite{Aharoni17}). (E)-(F) a schematic of the system shown in (A).  
(G)-(H) a schematic of the ``spherical Winkler'' model in (C). A thin sheet is attached to a sphere made of $N\gg 1$ radial harmonic springs, with rest length $R$ and spring constant $4\pi R^2 \Ksub/N$.}  
\end{figure*}

In the article, we aim to bridge the gap between the bending-dominated mechanics of Euler and strain-dominated geometry of Gauss through the ``Gauss-Euler {\it elastica}'' -- a proposed principle that generalizes of the variational rule (\ref{eq:Euler-var}) and characterizes the mechanical equilibrium of GIC problems.  We define GIC as the imposition of a smooth ``substrate'', whose {\emph{shape}} has 
a Gaussian curvature, $\Gsub({\bf x})$, onto a thin solid sheet or shell, whose ``target'' {\emph{metric}} is characterized by a different  curvature, 
$\Gtar({\bf x})$. 
In addition to $\Gsub\!- \!\Gtar \!\neq \!0$, a GIC problem is equipped with 
two dimensionless parameters, denoted $\chi^{-1}$ and $\epsilon^{-1}$, which characterize, respectively, the {\emph{confinement strength}}, and the {\emph{bendability}} of the confined solid.  In what follows, we will show how to define these parameters for other GIC problems ({\emph{e.g.}} Fig.~\ref{fig:1}B-F), but to begin, their meaning is best understood through the spherical stamping problem, where $\Gsub= R^{-2} > \Gtar=0$. Considering 
$t\ll \delta \ll W\ll R$, the confinement strength and bendability are conveniently defined as: 
\begin{equation}
\chi^{-1} = \Big(\tfrac{W^2/2R}{\delta}\Big)^2 \ \ \  ; \ \ \ 
\epsilon^{-1} = \Big( \tfrac{W^2/2R}{t}\Big)^2  \ ,
\label{eq:conf-bend-ss}
\end{equation} 
When $\delta \gtrsim W^2/R$ ({\emph{i.e.}} $\chi^{-1} \lesssim 1$), the wide gap requires no deflection and the sheet remains flat (i.e. no confinement).  If the gap is sufficiently small 
({\emph{i.e.}} $\chi^{-1} \gtrsim \epsilon^{-1} $), the stamp imposes perfect spherical shape, requiring a ``bare'' geometric strain $\varepsilonbare\!\sim\!(W/R)^2$, 
due to the elongation of its longitudes and shortening of latitudes. Our study focuses on the intermediate regime, $\epsilon^{-1}  \gg \chi^{-1} \gg 1$, in which the sheet shape is strongly sensitive to confinement, but can relieve the energetically costly strain down to a residual value, $\varepsilonres  \ll \varepsilonbare$, through deflections within the gap that are only penalized by their bending cost. We propose that these elastic deformations
are described by a variant of Eq.~(\ref{eq:Euler-var}):

\begin{gather} 
 {\rm Gauss-Euler \ {\emph{elastica}}:}  \nonumber \\
\delta \Ubend  = 0; \ 
 {\rm subject \ to}  \ \  {\Ustrain}/{\Ubend} \to  0 \ ,  
\label{eq:Gauss-Euler-var} 
\end{gather}   
where ``$\to$'' refers to the {\emph{doubly asymptotic}} limit, of high bendability and strong confinement, 
$\epsilon^{-1}\gg  \chi^{-1} \gg 1$.

Eq.~(\ref{eq:Gauss-Euler-var}) describes an energy minimization problem that is {\emph{dominated}} by bending, while being {\emph{constrained}} by the high cost of strain, much like the original {\it elastica} problem, but with two key and nontrivial distinctions. 
First,  the minimization principle, $\delta \Ubend =0$, means that {\emph{despite the presence of Gaussian curvature}}, the elastic energy is dominated by bending rather than strain; thus, Eq.~(\ref{eq:Gauss-Euler-var}) is closer 
to Euler's {\emph{elastica}} ~(\ref{eq:Euler-var}) than to the generic FvK problem~(\ref{eq:FvK-var}). 
Second, the suppression of strain that is expressed as an exact isometry constraint ($\Ustrain = 0$) in Euler's {\emph{elastica}}, is now imposed in a weaker, asymptotic manner ($\Ustrain/\Ubend \to 0$). Again, this latter distinction derives from the unavoidable distortion of a confined solid from its intrinsic metric due to a 
confining topography with $\Gsub \neq \Gtar$. We will show that the ratio $\Ustrain/\Ubend$, and 
the deformed shape of the confined solid,
are controlled by 
the confinement strength and bendability parameters, $\chi^{-1}$ and $\epsilon^{-1}$, 
yielding distinct types of energy minimizers 
at various sectors of the parameter regime, 
$\epsilon^{-1}\gg  \chi^{-1} \gg 1$.    

We commence 
with a simpler model 
of a 2D sheet attached to a ball of stiff, radial springs \cite{Grason13,Hohlfeld15,Bella17}. 
The relative simplicity of this ``minimal'' GIC problem enables a pedagogic exposition and an explicit derivation of the principle (\ref{eq:Gauss-Euler-var}), which we then employ to study 
spherical stamping. 

\section{Spherical Winkler foundation}
Consider a circular solid sheet of thickness $t$ and radius $W$, made of Hookean solid of Young's modulus $E$, with stretching and bending moduli, $Y=Et$ and $B \propto Yt^2$, respectively. The sheet is attached to a stiff spherical surface of radius $R$;  
deflections away from the substrate are penalized
by $N$ radial, uniformly spaced harmonic springs, each spring with constant $4\pi R^2 \Ksub/N$  
(Fig.1G-H). The energy density (per area), $u$, of the system is written schematically as:
\begin{equation}
u = \ustrain + \ubend + \usub  \ , 
\label{eq:energy-1}
\end{equation}
where: 
$\ustrain \!\sim\! Y \cdot (\text{strain})^2  \  ,  \ 
\ubend   \!\sim \! B \cdot (\text{curvature})^2 \ , $ 
and $\usub  \!\sim\!  \Ksub \cdot (\text{deflection})^2 \ .$
The strain and curvature tensors, $\varepsilon_{ij}$ and $\kappa_{ij}$, respectively, are given in terms of the
in-plane displacements, $\rmu_r,\rmu_{\theta}$, and normal deflection $\zeta$ of the sheet from its rest, planar state 
({\emph{Appendix}}).  
Here, the two dimensionless groups that quantify the 
strength of confinement and bendability of the sheet are defined as \cite{Grason13,Hohlfeld15}: 
\begin{equation}
\chi^{-1} = \frac{\Ksub R^2}{Y} \ \ ; \ \ \epsilon^{-1} = \frac{YW^4}{BR^2} \sim \Big(\frac{W^2}{tR}\Big)^2 \  .   
\label{eq:Winkler-parameters}
\end{equation}    
The physical meaning of these parameters can be grasped by considering 
an ideal, axisymmetric deformation of the system.
If $\Ksub$ is very small, the sheet remains nearly planar and the energetic cost of flattening the substrate beneath it 
is $\usub \sim \Ksub (\tfrac{W^2}{R})^2$, whereas if $\Ksub$ is large the substrate is barely deformed and the sheet conforms to its spherical shape  
with characteristic energies, $\ubend \sim B/R^2$ and $\ustrain \sim Y\varepsilonbare^2$ (where $\varepsilonbare \sim (W/R)^2$). In this light, the dimensionless parameters in Eq.~(\ref{eq:Winkler-parameters}) are seen as 
the ratios 
$\chi^{-1} \sim \tfrac{\usub}{\ustrain}$, 
$\epsilon^{-1} \sim \tfrac{\ustrain}{\ubend}$.  
Here we focus on the regime: 
\begin{equation}
\epsilon^{-1} \gg \chi^{-1} \gg 1 \ \  \ \ \Longrightarrow  \  \chi \ll 1 \ , \ (\epsilon/\chi) \ll 1\  \ , 
\label{eq:regime-1}
\end{equation}
in which $\Ksub$ is large, and the substrate closely retains its shape to avoid a high energy $\usub$. However, 
rather than conforming axisymmetrically to the substrate, which  
would generate a bare strain 
and an energy density, $\uaxi \approx \ustrain \sim Y\varepsilonbare^2$,  
the high bendability of the sheet enables a substantial suppression of strain, 
down to a residual level,  $\varepsilonres$, via formation of small-wavelength wrinkles that ``absorb'' the excess length of compressed latitudes at the expense of only slight deviation of the substrate from its spherical shape. 

\begin{figure*}
\center
\includegraphics[width=1.0\textwidth]{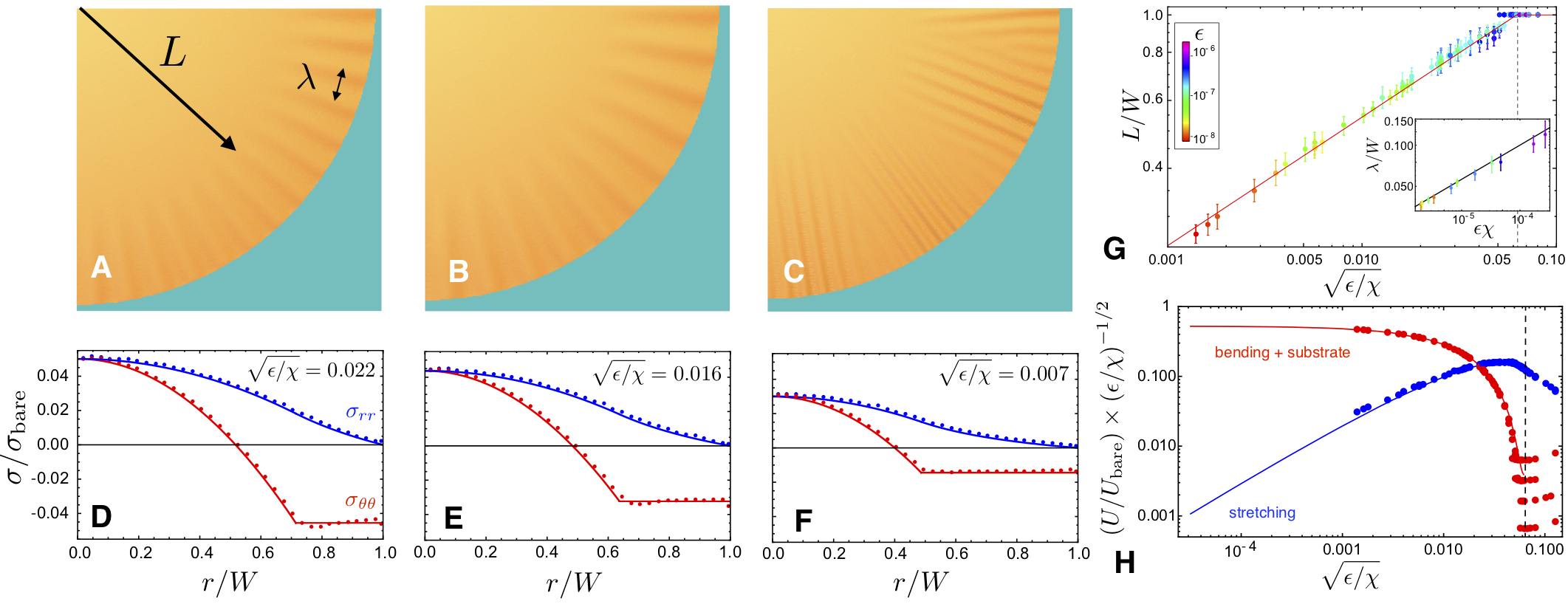} 
\caption{\label{fig:2}  (A-C) top views of simulated circular sheets 
(gold) bound to spherical Winkler substrate (blue), with corresponding profiles of radial ($\sigma_{rr}$, blue) and hoop ($\sigma_{\theta \theta}$, red) shown in (D-F): circles show stress computed from simulations and solid curves show ITFT predictions ({\emph{Appendix}}). 
Stresses are normalized by the bare stress scale, 
$Y \varepsilonbare = Y(W/R)^2$. (G) comparison of observed values of $L$ (radius of unwrinkled core) and $\lambda$ (wrinkle wavelength, measured at $r\!=\! W$, inset), with ITFT predictions (Eq.~\ref{eq:wave-length} and {\emph{Appendix}}), 
plotted as functions of  $\epsilon/\chi$, the ratio of the inverse bendability ($\epsilon$) and confinement strength ($\chi$), Eq.~(\ref{eq:Winkler-parameters}). 
Colors indicate the value of $\epsilon$, 
and vertical dashed line denotes the wrinkling threshold. 
(H) plots of the energy content in bending $+$ substrate deformation (red) and strain energy (blue) for simulated sheets, showing that the ratio 
between strain and bending energies vanishes as $\epsilon/\chi \to 0$, as invoked by 
the Gauss-Euler {\it elastica} (Eq.~\ref{eq:Gauss-Euler-var}). Energies are normalized by the bare energy  
of the unwrinkled state 
($\pi  Y W^6/( 384 R^2) \! \sim\! Y \varepsilonbare^2 W^2$); ITFT predictions ({\emph{Appendix}}) are shown in solid curves.}
\end{figure*}

The deformed shape and the residual strain can be found via a numerical simulation of the Seung-Nelson model of a 2D elastic sheet bound to spherical Winkler foundation~\cite{NelsonSeung,Rastko}, whose discrete elements capture the full geometrical nonlinearity underlying the FvK energy ({\emph{Appendix}}).  
In our simulations (Fig.~2) we explore the relevant parameter regime (\ref{eq:regime-1}) by fixing the confinement strength at a large value, $\chi^{-1} \simeq 39, 78, 156$ or $340$, and gradually increasing the bendability, $\epsilon^{-1}$ (from $2 \times 10^{5} $ to $10^{8}$) by reducing the
sheet's thickness. 
As $\epsilon^{-1}$ exceeds a critical value (dashed vertical in Fig. 2G,2H), 
radial wrinkles emerge in an annular zone, $L<r<W$. Increasing $\epsilon^{-1}$ further, 
the wrinkles cover a larger and larger portion of the sheet, and their characteristic wavelength, $\lambda$, becomes smaller (Fig.~2G). We find that: 
\begin{equation}
\lambda/W \approx 2\pi(\chi \epsilon )^{1/4}  \ \ ; \ \ {L}/{W} \simeq 2^{4/3}  (\epsilon/\chi)^{1/6} 
\label{eq:scaling-11}
\end{equation}
Analyzing the radial and hoop components of the in-plane stress tensor, $\srr(r), \sqq(r)$ (which are linear combinations of the respective components of the strain tensor, see {\emph{Appendix}}), we find that 
unlike the axisymmetric case, for which $\srraxi \sim \sqqaxi \sim Y \varepsilonbare^2$, here: 
 \begin{equation}
 \srr \sim \sqq \sim  Y\cdot \varepsilonres \ \ , \text{where:} \ \ \varepsilonres \approx \sqrt{B\Ksub}/Y  \ , 
 \label{eq:stress-relax}
 \end{equation}
(Figs.~2D-2F), such that $\varepsilonres/\varepsilonbare \sim \sqrt{\epsilon/\chi} \to 0$ in the limit (\ref{eq:regime-1}).
Let us recall that uniaxial compression of  supported sheets yields wrinkles with wavelength $\lambda$ of the form (\ref{eq:scaling-11}), reflecting suppression of compressive stress (here $\sqq$) down to residual level $\sim \!- \sqrt{B\Ksub}$ (obtained by balancing sheet bending and substrate stiffness)\cite{Bowden98,Cerda03}. 
Figures.~2D-2F show that not only the compressive hoop stress, $\sqq$, but also the {\emph{purely tensile}}
radial stress, $\srr$, 
scales as $\sqrt{B\Ksub}$.

Further analysis of simulations reveals that the wrinkle-assisted suppression of stress (\ref{eq:stress-relax}) reshuffles the energetic hierarchy in the asymptotic limit (\ref{eq:regime-1}). Fig.~2H shows that the strain energy is dominant at the vicinity of threshold, but becomes negligible in comparison to the bending and substrate energies in the limit (\ref{eq:regime-1}). 
This reversal of the energetic hierarchy can be understood through estimates of respective energy densities which, at first pass, neglect spatial variation. Since Eq.~(\ref{eq:stress-relax}) shows that all components of strain scale as $\varepsilonres$,  
we have that $\ustrain  \sim Y \varepsilonres^2$, while $\ubend \sim B (f/\lambda^2)^2$, where $f$ is the characteristic amplitude of wrinkles, the wavelength, $\lambda$, is given by (\ref{eq:scaling-11}), and $\usub \sim \Ksub f^2$. 
To estimate $f$, we employ
a ``slaving'' condition that links the wrinkle amplitude and wavelength, $(f/\lambda)^2 \sim (W/R)^2$, expressing the fact that the excess arclength ``wasted'' by azimuthal wrinkles matches the inward radial displacement of latitudes on the substrate \cite{Hohlfeld15,Bella17}. Together with Eq.~(\ref{eq:scaling-11}) we thus obtain: 
\begin{equation}
\ubend \sim \usub \sim \uaxi   \sqrt{\epsilon/\chi}  \ \ ; \ \  
\ustrain/\ubend \sim \sqrt{\epsilon/\chi} \ . 
\label{eq:energy-ratio-1}
\end{equation}
This result 
exhibits two critical features. First, the elastic energy is asymptotically vanishing in comparison to the bare energy, $\uaxi \sim Y\varepsilonbare^2$, required for perfectly conforming the flat sheet to a sphere. 
Second, the energetic hierarchy is reversed in comparison to the bare deformation -- 
being 
governed by bending 
(and substrate deformation) rather than strain. 
Taken together, these two features manifest the Gauss-Euler principle (\ref{eq:Gauss-Euler-var}). Notwithstanding the fact that 
imposing Gaussian curvature generates {\emph{some}} strain in the sheet, 
its energetic cost 
is negligible and the energy-minimizing state is found by minimizing bending energy (for this case, in balance with the substrate) over a family of ``asymptotically isometric'' configurations, in an analogous manner to Euler's {\emph{elastica}} (\ref{eq:Euler-var}).

\section{Inverted tension field theory (ITFT)}
Motivated by the numerical results, and by classical tension field theory (TFT), which describes an asymptotic, {\emph{compression-free}} stress field in confinement problems that are governed by tensile boundary loads  \cite{Mansfield}, we introduce here an   
``inverted'' tension field theory (ITFT) that describes an asymptotic, {\emph{strain-free}} stress field, in confinement problems governed by geometrical incompatibility. For the spherical Winkler problem, this theory is formulated as a {\emph{doubly-asymptotic}} expansion of FvK equations in 
$\chi $ and $\epsilon$, around the singular 
limit (\ref{eq:regime-1}), to which we refer below by the symbol ``$\to$''. The succinct exposition below is supported by a detailed analysis of the displacement field, force balance equations and energy balance in {\emph{Appendix}}.

{\it 1. Hoop confinement}: Elimination of radial strain, namely, preserving length of longitudes, requires a finite radial displacement: $\rmu_r(r) \to -\tfrac{1}{6} \tfrac{r^3}{R^2}$. This follows directly from the condition $\varepsilon_{rr} = \partial_r\rmu_r + \tfrac{1}{2} (\tfrac{\partial \zeta}{\partial r})^2\ \to 0$, where 
$\zeta_{sph}(r) \approx - r^2/2R$. Consequently, 
the projection of sheet's latitudes onto the sphere 
is contracted by: $\tDelta(r) \equiv -\rmu_r/r \to  \tfrac{1}{6} (\tfrac{r}{R})^2$. 

{\it 2. Wrinkle wavelength and hoop stress}: Considering each latitude as an elastic hoop attached to a substrate of stiffness $\Ksub$ under confinement 
$\tDelta(r)$, wrinkles relax the bare compressive strain, $\tDelta(r)$, namely: 
$\zeta(r,\theta) \to \zeta_{sph}(r) + f(r)\cos(r \theta/\lambda)$, such that the amplitude, $f$, and wavelength, $\lambda$, satisfy the aforementioned ``slaving'' condition $(\tfrac{\pi f}{\lambda})^2\! \!\to\! \!\tDelta(r)$ \cite{Davidovitch11}. Minimizing bending and substrate energies one obtains:   
\begin{equation}
\lambda \approx 2\pi (B/\Ksub)^{1/4}  = W (\epsilon \chi)^{1/4} \ ; \ \sqq \approx  -2B/\lambda^2  \ .  
\label{eq:wave-length}
\end{equation}

{\it 3. Radial stress}: 
Turning now to force balance in the radial direction, we address the $2^{nd}$ FvK equation:
\begin{gather}
\partial_r(r\srr) - \sqq = 0 \label{eq:radial-force-balance} \\
 \Longrightarrow \srr \to 2\sqrt{B\Ksub} (-1 + W/r) \ , \label{eq:radial-stress}
\end{gather} 
where we employed $\sqq$ 
from Eq.~(\ref{eq:wave-length}) as a {\emph{non-homogeneous}} source for $\srr$ along with the free boundary condition, $\srr|_{r\!=\!W} \!=\!0$. Notably, Eq.~(\ref{eq:radial-stress}) manifests the counter-intuitive concept of ``bending-induced'' tension {\emph{along}} wrinkles, which 
was envisioned already by Hure {\emph{et al.}} \cite{Hure12}. 
 
{\it 4. Strained core}: The spatial divergence of radial stress, $\srr \sim r^{-1}$ in (\ref{eq:radial-stress}), must be alleviated by 
a strained, unwrinkled core, $0\!<\!r\!<\!L$, whose radius vanishes asymptotically as $\chi,\epsilon \to 0$. The characteristic strain in this core is $ (L/R)^2$, yielding a radial (and similarly, hoop) tension, $\srr \!\sim\! Y (L/R)^2$. The continuity of $\srr$ at $r=L$  is required by radial force balance and yields the scaling relation: $L/W \!\sim\! (\epsilon/\chi)^{1/6}$, 
 in accord with our simulations (\ref{eq:scaling-11}). Notably, the core-assisted regularization of radial stress (\ref{eq:radial-stress}) modifies our estimate of the strain energy (\ref{eq:energy-ratio-1}) by only a logarithmic factor, $\log(W/L) \!\sim\! \log(\epsilon/\chi)$.

{\it 5. Vanishing shear}: 
The above arguments 
explain why the radial and hoop stresses vanish
as $\chi,\epsilon \to 0$. However, an asymptotically-isometric configuration 
requires also the vanishing of shear stress, $\srq \to 0$,
which is accommodated by way of small oscillations of radial displacement in registry with wrinkles, 
$\rmu_r   \approx - r\tDelta
+ (fr/R) \cos ( r \theta / \lambda)$ 
\cite{Paulsen16}. The energetic cost of these ``shear cancelling'' radial oscillations leads to curvature-induced stiffness, $ \Kcurv \!= \! Y/R^2$, 
that suppresses further the wrinkle amplitude.  Taken together with the 
substrate stiffness, this leads to a generalized version of Eq.~(\ref{eq:wave-length}): 
\begin{equation}
\lambda \!=\! 2\pi({B}/{\Keff})^{1/4}  \ ;  \  \Keff \!= \!\Ksub \!+\! \Kcurv  \ .  
\label{eq:local-lambda}
\end{equation}   
(Since $\srr \ll \Ksub/W^2$, another contribution to $\Keff$, due to tension along wrinkles \cite{Paulsen16}, is negligible here). 
Since $\Kcurv/\Ksub = \chi \ll 1$, curvature-induced stiffness is also negligible here. 
However, we will show below that the elastic cost of shear suppression 
is critical for other GIC problems.
  
The  solution of our ITFT equations yields quantitative expressions for the wavelength (Fig.~2G) and residual stresses (solid curves in Figs.~2D-2F), as well as the energies, 
$\Ubend, \Usub$, and $\Ustrain$ (Fig.~2H), in excellent agreement with simulations. This substantiates the Gauss-Euler {\emph{elastica}} (\ref{eq:Gauss-Euler-var}) and the ITFT equations, whose polar representation is Eqs.~(\ref{eq:wave-length},\ref{eq:local-lambda},\ref{eq:radial-force-balance}), as valuable tools for solving GIC problems.

\section{ITFT versus TFT}
We elucidate how ITFT embodies the Gauss-Euler {\emph{elastica}} (\ref{eq:Gauss-Euler-var}) by comparing the radial force balance equation and 
radial stress, Eqs.~(\ref{eq:radial-force-balance},\ref{eq:radial-stress}), with their 
counterparts
in two standard studies of a sheet attached to a spherical substrate (Fig.~3A). 

The ``bare'', axisymmetric (unwrinkled) state results from solving Eq.~(\ref{eq:radial-force-balance}) by expressing the stress tensor through the axisymmetric displacement field (Fig.~3A, top row) \cite{Grason13}, yielding 
a $2^{nd}$ order equation for $\rmu_r(r)$. The consequent mean strain energy is $ \langle \ustrain \rangle = \Ustrain/\pi W^2 \sim Y \varepsilonbare^2$.    

A TFT analysis of this problem (Fig.~3A, middle row) \cite{Grason13,Hohlfeld15} is suitable when a tensile load, $\gamma=\srr |_{r\!=\!W}$, is sufficiently smaller than $Y\varepsilonbare$, 
yet much larger than the residual hoop compression in a wrinkled state, $|\sqq| \sim 2\sqrt{\Ksub B}$ (\ref{eq:wave-length}). Under such conditions, Eq.~(\ref{eq:radial-force-balance}) becomes a $1^{st}$ order {\emph{homogenous equation}} for the radial stress, and the asymptotic stress tensor has a single non-vanishing component: $\tiny{\sigma_{ij} \!\to\! \gamma   
\left( \begin{array}{cc}
W/r & 0  \\
0 & 0  \end{array} \right)} $, which underlies the dominant part of the energy.    

In contrast, 
the ITFT stress yields a 
subdominant energy
($\ustrain \! \ll \! \ubend$). 
Equation~(\ref{eq:radial-force-balance}) becomes 
an {\emph{inhomogeneous}} equation for $\srr$, whose source ($\sqq$) is set {\emph{a priori}} by minimization of the bending energy,  
disregarding the explicit contribution of $\ustrain$ to the total energy.  
The asymptotic stress field has the form $\tiny{\sigma_{ij} \!\to\! 2\sqrt{\Ksub B}   
\left( \begin{array}{cc}
-1+W/r & 0  \\
0 & -1  \end{array} \right)} $, acting as a {\emph{tensorial Lagrange multiplier}} that 
generalizes the 
scalar Lagrange multiplier of Euler {\emph{elastica}}.

\begin{figure}
\center
\includegraphics[width=0.5\textwidth]{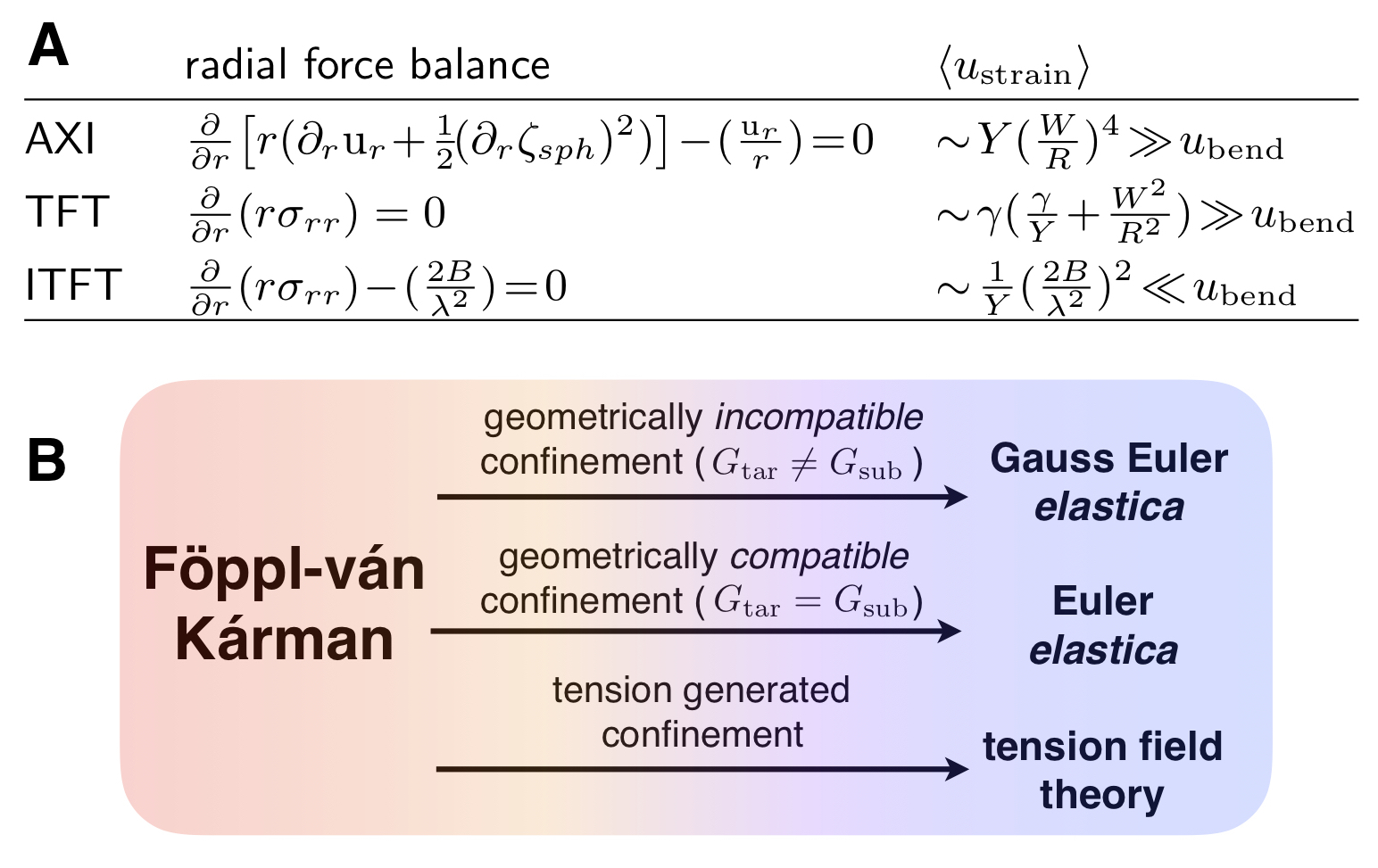} 
\caption{\label{fig:3} (A) three incarnations of radial force balance (\ref{eq:radial-force-balance}) for a spherically confined sheet, where $\langle \ustrain\rangle = \Ustrain/\pi W^2$. Standard TFT addresses a tensile load $\gamma$ at the edge, $r=W$, 
whose work is included in $\langle \ustrain\rangle$ (a logarithmic factor, $\log(W/L)$, is neglected from the energies of TFT and ITFT). For simplicity, the axisymmetric equation is written for zero Poisson ratio. (B) the mechanical equilibrium of a thin elastic solid is described by the FvK energy (\ref{eq:FvK-var}), and requires simultaneous minimization of straining and bending energies. As the bendability of the confined solid increases (arrows direction), the problem is described by an effective rule, depending on the conditions that generate confinement.}     
\end{figure}

\section{The spherical stamping problem} 
Spherical stamping differs from the Winkler foundation problem in that there is no energetic penalty for deforming the spherical ``substrate''. 
Instead, the energy consists only of bending and straining the sheet, subjected to the constraint that its deflection from the sphere is bounded: $|\zeta(r,\theta) - \zeta_{sph}(r)| \leq \delta$. A qualitative understanding of this problem can be obtained by assuming that wrinkles fill the gap, such that the slaving condition, $(\pi  f/\lambda)^2 \!= \! \tDelta(r) \sim (r/R)^2$, implies a wrinkle wavelength $\lambda \sim \delta$ 
(up to $r$-dependent pre-factors). 
The ``local $\lambda$ law'' (Eq.~\ref{eq:wave-length}) then suggests that 
the gap effect is akin to a spherical Winkler substrate: $\Ksub(r) \!\sim\! B/\delta^4$ ({\emph{Appendix}})~ \cite{Paulsen16}. 
The ITFT equations~(\ref{eq:wave-length},\ref{eq:radial-force-balance}) thus yield $|\sqq| \!\sim\! \srr\! \sim \!B/\delta^2$. 
Using the results of the preceding analysis and recalling the dimensionless groups defined in Eq. (\ref{eq:conf-bend-ss}) we readily find that for a given ratio $\epsilon/\chi = (\delta/t)^2$, 
radial wrinkles cover the sheet barring a core of radius
$L/W \sim (t/\delta)^{2/3}$. 
Evaluating the hoop and radial stresses 
and the various energies, one finds that: 
$\Ubend/\Ubare \sim (t/\delta)^2$ and $\Ustrain/\Ubend \sim (t/\delta)^2$. Hence, the deformation satisfies the Gauss-Euler {\emph{elastica}} (\ref{eq:Gauss-Euler-var}) in the doubly asymptotic limit, $\chi,\epsilon/\chi \to 0$.

\begin{figure*} \center \includegraphics[width=0.95\textwidth]{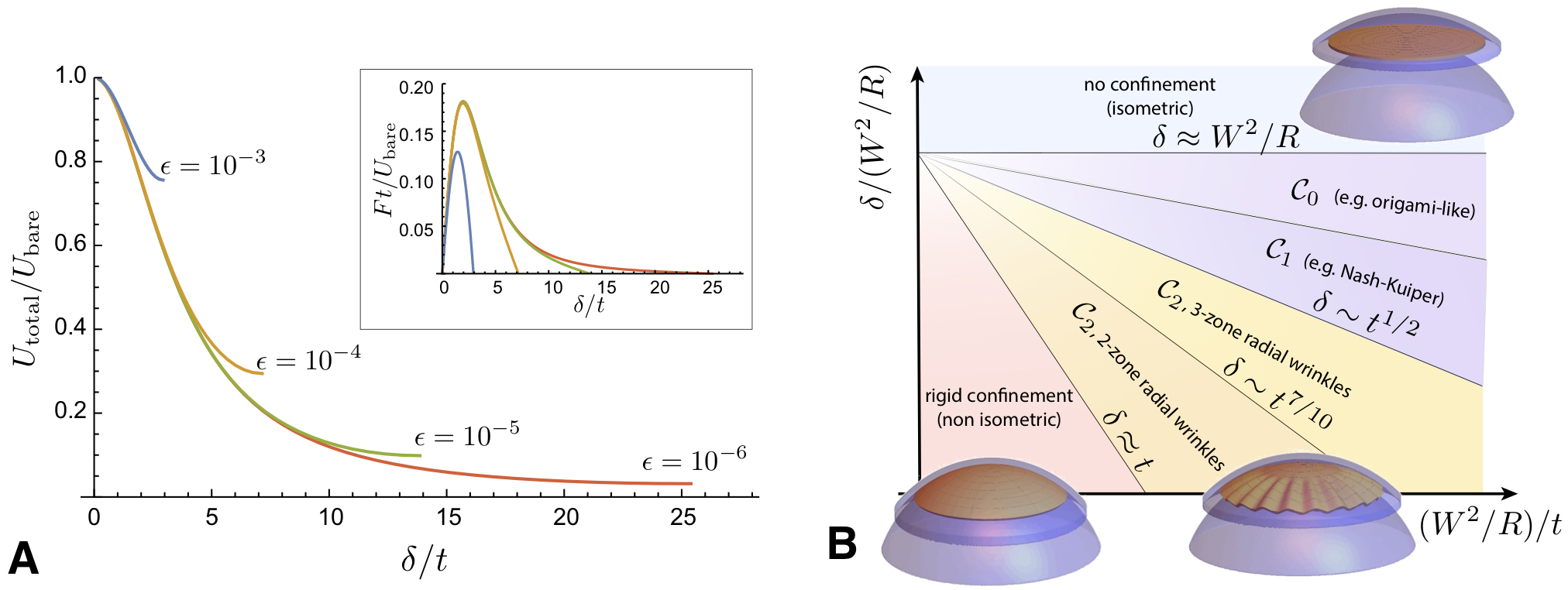}  
\caption{\label{fig:4} ITFT predictions for the spherical stamping problem. 
(A) the predicted total energy, $U_{\rm tot}$, and force, $F = 2 \partial_\delta U_{\rm tot}$ 
(inset), plotted {\emph{vs.}} the (normalized) gap separation $\delta/t$, for a range of inverse bendability parameters.  
Energies are normalized by the bare energy (of axisymmetric state), 
$U_{\rm bare} = \pi  Y W^6/( 384 R^2)$. Curves terminate at the point when wrinkles break contact with stamp, namely, $L_1 \to W$. 
(B) a schematic phase diagram of the spherical stamping problem, spanned by gap height $\delta$ and (inverse) sheet thickness $t^{-1}$, 
with the predicted regimes of 2-zone or 3-zone radial wrinkle patterns highlighted in orange/yellow.  } \end{figure*}

Note that if $\delta\! \gtrsim \!\sqrt{Rt}$ ($\chi\! \ll\! \epsilon^{1/2}\!\ll\! 1$), the gap-induced stiffness, $\Ksub\!=\! B/\delta^4$, falls below 
$\Kcurv \!=\! Y/R^2$. In this regime Eq.~(\ref{eq:local-lambda}) restricts the wavelength to: $\lambda\!\! \sim \!\!\sqrt{Rt}$; consequently, wrinkles do not fill the gap ({\emph{i.e.}} $f \!\!<\!\!\delta$). 
This reflects the energetic cost of supressing shear strain when wrinkling about a curved shape; 
wrinkles with amplitude larger than $\sim\!\sqrt{Rt}$ would not satisfy $\ustrain\ll \ubend$, and thus, do not constitute an asymptotic solution of the Gauss-Euler {\emph{elastica}}~(\ref{eq:Gauss-Euler-var}).  

In order to obtain the 
actual ITFT solution for the spherical stamping problem, one has to consider a ``3-zone'' pattern: a ``gap limited'' zone ($L_1\!\!< \!\!r \!\!< \!\!W$), where the amplitude $f(r)=\delta$, 
maximizing deflection at the vicinity of the edge where confinement is strong; a ``curvature-limited'' zone ($L_2\!\!<\!\! r\!\!<\!\!L_1$) -- where the wavelength $\lambda \approx \sqrt{Rt}$ and $0\!\!<\!\!f(r)\!\!<\!\!\delta$, so that wrinkles do not fill the gap; and finally a strained, unwrinkled core at $r\!\!<\!\!L_2$ 
(the geometry and distribution of stamp forces are described in
{\emph{Appendix}}). 
A central result of this analysis 
is the emergence of two ``sub-phases'' of ITFT (see Fig.~4B): 
For $\epsilon \! \lesssim \!\chi\! <\! \epsilon^{7/10}$ ($t \!\lesssim \!\delta \!\lesssim \!t^{7/10}$), the pattern comprises only a gap-limited zone, which terminates sharply at an unwrinkled core ($L_1 = L_2\sim W (t/\delta)^{2/3}$ in the above notations); the intervening, curvature-limited zone appears when $\chi \sim \epsilon^{7/10} (\delta \sim t^{7/10})$, and expands (with $L_2\sim W (t /\delta)^{2/3}$ and  $L_1 \sim \delta (R/t)^{1/2}$) upon increasing the gap until it reaches the outer edge at $\chi \sim \epsilon^{1/2}$ ($\delta \sim t^{1/2})$, at which stage a wrinkled state loses contact with the confining shells. 
Figure~4A presents the energy $U$ of our ITFT solution (normalized by $\Ubare = \pi Y W^6/ 384R^4$), and the related stamping force $F = \partial U/\partial \delta$, as a function of $\delta/t$, for some values of the inverse bendability, $\epsilon$, exhibiting two noteworthy features. First, each plot terminates at a finite value, $(\delta/t)_{max} \!\sim \!\epsilon^{-1/2}$, at which the wrinkle pattern is fully curvature-limited and detaches from the confining shells (signified by a vanishing stamping force). Second, as bendability increases, the energy relaxation becomes more efficient -- the residual energy, $U_{\rm tot}$ at  $(\delta/t)_{max}$ scales as 
$\epsilon^{1/2} \ \Ubare$. 

\section{Discussion}
We introduced the Gauss-Euler {\emph{elastica}} (\ref{eq:Gauss-Euler-var}) for GIC problems -- stating that bending is minimized subject to negligibility of strain energy.
We demonstrated the applicability of this rule in two variants of a spherically-confined sheet through ITFT -- an asymptotic expansion of FvK equations around a singular, strain-free limit. 
We showed that the ITFT may be regarded an extension 
of Euler {\emph{elastica}}, where the stress acts as a tensorial, spatially-varying Lagrange multiplier that enforces asymptotic isometry. 
The schematic, Fig.~3B, shows how the highly nonlinear FvK problem 
for a confined sheet (\ref{eq:FvK-var}) recasts distinct simplifications
as the sheet thickness becomes small: Euler {\emph{elastica}} is applicable for developable deformations (no tension, no Gaussian curvature); TFT is applicable in the presence of tensile loads whose work dwarfs bending energy; ITFT applies for GIC problems, where non-developable deformations emerge in the absence of tensile loads.

The GIC problem we addressed -- confinement of a planar sheet onto a spherical topography -- corresponds to $\Gtar \!\!< \!\!\Gsub$, where confinement is preferentially azimuthal and the emerging wrinkle pattern is splayed. The GIC problem shown in Fig.~1B is characterized by $\Gtar \!\! >\! \! \Gsub$; there, confinement is preferentially radial and wrinkles are bent. We expect that the Gauss-Euler {\emph{elastica}} (\ref{eq:Gauss-Euler-var}) is valid also for such problems, and even for more complex GIC problems 
that result in a mix of splay and bend of wrinkle textures (Fig.~1D).

Our analysis assumes that the minimal FvK energy (\ref{eq:FvK-var}) can be safely evaluated through the upper bound obtained by solving the ITFT equations for smooth, periodic deformations ({\emph{Appendix}). Notwithstanding the agreement between our simulations (Fig.~\ref{fig:2}) and solutions of the ITFT equation~(\ref{eq:wave-length},\ref{eq:local-lambda},\ref{eq:radial-force-balance}), we emphasize that the validity of the rule~(\ref{eq:Gauss-Euler-var}) hinges on the validity of this central assumption, which awaits a rigorous mathematical proof. Furthermore, while our findings support the validity of Eq.~(\ref{eq:Gauss-Euler-var}) for the class of {\emph{strong}} confinement problems addressed here, identified by finite, thickness-independent incompatibility,  
$|\Gtar-\Gsub| \neq 0$, we suspect that it fails for a class of ``weak confinement" problems for which the mismatch $|\Gtar-\Gsub|$ may eventually vanish with $t$.  
Well known examples for this latter class of problems are the biaxial confinement of a stiff skin attached to a (planar) compliant substrate \cite{Bowden98,Audoly08,Nguyen13}, and the contact of curved liquid surfaces 
with a planar sheet \cite{King12}. The complexity of such problems can be anticipated by considering the two problems 
we addressed here in the parameter regimes complementary to (\ref{eq:regime-1}), namely, $\epsilon \ll 1, \chi/\epsilon \ll 1$. In the spherical Winkler problem, this corresponds to the regime 
$\chi^{-1} \!\ll\! 1$, where the substrate is very soft and can deform substantially -- flattening uniformly to reduce sheet's strain and thereby developing stress-focusing, crumpled zones \cite{King12}, or alternatively, enabling the formation of localized 
folds \cite{Pocivavsek08,Diamant11,Paulsen17a}.

For the spherical stamping problem, a parameter regime that corresponds to ``weak confinement'' is, $\delta \!\gg\! \sqrt{Rt}$, where 
radial wrinkles predicted by ITFT no longer fill the gap, and another deformation is required to relax the residual energy, $\sqrt{t/\delta}\cdot \Ubare$, stored in a wrinkled state (Fig.~4A). To wit, let us consider a very weak confinement, namely, $\delta \sim t^\beta$, where $0\!<\!\beta\!\ll\! 1$ (light purple in Fig.~4B); here, the sheet barely touches the confining shells, and the problem seems qualitatively similar to confining a sheet in a ball whose radius is only slightly smaller than its own \cite{Aharoni10}. There,  
the deformation is governed by a few ``stretching ridges'' 
that 
separate the sheet into stress-free facets \cite{Witten07}. 
Such a deformation is distinguished from a wrinkled state in two intimately-related aspects. First, the ridges, which dominate the energy, are characterized by a balance of bending and strain \cite{Witten07}, thereby obeying the general FvK rule (\ref{eq:FvK-var}), but not the restrictive version invoked in Gauss-Euler {\emph{elastica}} (\ref{eq:Gauss-Euler-var}). Second, in contrast to wrinkle patterns, where curvature oscillates throughout the sheet, a network of ridges localizes curvature in ``boundary layers" (whose width vanishes as $t\to 0$) and consequently an asymptotic divergence of curvature and discontinuity of slope between the two sides of each ridge, similarly to Origami constructions. 


The plausibility of origami-like deformations under weak confinement suggests yet another possibility of a smoother but nevertheless non-wrinkly deformation: curvature, rather than slope, is spatially non-uniform, such that only the curvature's derivative becomes localized in a network of ridges, and diverges as $t\to 0$ (heavy purple in Fig.~\ref{fig:4}B).  
Such a nontrivial pattern is suggested by Nash embedding theorem \cite{Nash}, whose relevance for the deformations of solid sheets has very recently begun to be explored \cite{Gemmer16}. Whether such Nash mappings do characterize an asymptotically-isometric response of confined solids, and whether their regularized versions (for small, but finite $\delta$ and $t$) are subjected to the Gauss-Euler {\emph{elastica}} (\ref{eq:Gauss-Euler-var}), are fascinating questions that we hope will inspire future studies.

\begin{acknowledgments}
We thank P. Bella, R.V. Kohn, and N. Menon for many discussions, B. Roman, J. Bico and J. Hure for discussing with us their experiment and model \cite{Hure12}, R. Sknepnek for his assistance with numerical simulations, D. Vella, D. O'Kiely, J. Bico and J. Paulsen for useful comments, and participants of the program ``Geometry and elasticity of 2D soft matter'', KITP Santa-Barbara 2016. 
This work was funded by the ACS Petroleum Research Fund Grant No. 54513-ND  and NSF Award CBET-14-38425. Simulations were performed using the UMass Cluster at the Massachusetts Green High Performance Computing Center. 
\end{acknowledgments}

\appendix

\section{A sheet on a slightly deformable sphere}

Here we describe in detail the ``spherical Winkler'' model presented in the main text. 
%




We consider a circular elastic film of radius $W$ attached to a ``spherical Winkler'' substrate, or equivalently  
a ball made of radial, non-interacting harmonic springs of stiffness $4\pi \Ksub R^2/N$ and rest length $R \gg W$ (Fig.~1 of main text). We assume the sheet to be perfectly attached to the substrate ({\emph{i.e.}} an infinite adhesion energy); however, the force exerted by the deformed substrate on the sheet is purely normal to the sheet's midplane, with no tangential component (such that the sheet slides freely on the substrate surface).   This model captures the essential mechanics of adhesion onto a curved isotropic solid, here simplified by assuming that the substrate response is local and without tangential forces. 
This ``spherical Winkler'' model was described in Sec.~II of \cite{Hohlfeld15}; here we will briefly review its main features, focusing on the case of a sheet with {\emph{stress-free boundary.}}
       
%
Since $W\ll R$, the displacement can be expressed through the Monge parametrization: 
\begin{equation}
\mathbf{u}(r,\q) = \ru_r(r,\q)\mathbf{\hat r} +  \ru_\q(r,\q)\boldsymbol{\hat\q} +  \zeta(r,\q) \boldsymbol{\hat z} \   , 
\label{eq:displacement}
\end{equation}
with the components of the strain tensor, $\boldsymbol\varepsilon$, given by 
\begin{subequations} \label{eq:strains}
\begin{gather}
\varepsilon_{rr} = \partial_r \ru_r + \tfrac{1}{2} (\partial_r\zeta)^2\ , \label{eq:strain-radial-1} \\
\varepsilon_{\q\q} = \tfrac{1}{r} \partial_\q \ru_\q + \tfrac1r \ru_r + \tfrac{1}{2r^2} (\partial_\q \zeta)^2 \ ,\label{eq:hoopstrain}\\
\varepsilon_{r\theta} = \varepsilon_{\theta r} = \tfrac12 \left( \tfrac1r \partial_\q \ru_r +  \partial_r \ru_\q + \tfrac1r \partial_r\zeta\partial_\q \zeta\right)\ ,  \label{eq:strain-shear-1}
\end{gather}
\end{subequations}
The stress in the sheet is given by the Hookean relationship \cite{LL86, timoshenko70, Mansfield}: 
\begin{subequations} \label{eq:stresses}
\begin{gather}
\sigma_{rr} = \frac{Y}{1-\Lambda^2} \left(\varepsilon_{rr} + \Lambda \varepsilon_{\q\q} \right)\ \label{eq:stresses-radial} , \\
\sigma_{\q\q} = \frac{Y}{1-\Lambda^2} \left( \varepsilon_{\q\q} + \Lambda \varepsilon_{rr} \right)\ , \\
\sigma_{r\theta} = \frac{Y}{1+\Lambda} \varepsilon_{r\theta} \ , 
\end{gather}
\end{subequations}
where $\Lambda$ the Poisson ratio of the sheet. 
Furthermore, anticipating the shape $\zeta(r,\theta)$ to be characterized by small slopes (even in the wrinkled zones), the curvature tensor $\kappa_{ij}$ can be approximated as: 
\begin{equation}
\kappa_{rr} = \partial^2_{rr}\zeta \ \ ; \ \ \kappa_{\theta\theta}  = \tfrac{1}{r} \partial_r \zeta + \tfrac{1}{r^2} \partial^2_{\theta\theta} \zeta \ \ ; \ \ \kappa_{r\theta} = \frac{2}{r} \partial^2_{r\theta} \zeta 
\label{eq:disp-curvature}
\end{equation} 
The energy ($U$) is conveniently expressed through its areal density ($U=\int dA~u$): 
\begin{subequations} \label{eq:FvKenergy}
\begin{equation}
u = \uFvK + \usub \ , 
\label{eq:EnWinModel}
\end{equation}
where $u_{\rm FvK} = u_{\rm strain} + u_{\rm bend}$ , and: 
\begin{gather}
\usub  = \frac{\Ksub}{2}  \big(\zeta+r^2/2R\big)^2 \ ,  \nonumber  \\
u_{\rm strain} = \tfrac{1}{2} \sigma_{ij} \varepsilon_{ij}  \ \ ; \ \ 
u_{\rm bend} = \frac{B}{2} Tr\boldsymbol(\kappa)^2 
\label{eq:EnWindefine}
\end{gather}
\end{subequations}
(Eq.~5 of main text). In Eq.~(\ref{eq:EnWindefine}) we once again employed the inequality, $W\ll R$, approximating the rest (spherical) state of the substrate as $\zeta_{sph}(r) \approx -r^2/(2 R)$. Furthermore, we will see later that the bending energy is  governed by the rapid, azimuthal undulations of the shape, and hence can be well approximated as $u_{\rm bend} \approx \tfrac{B}{2}  r^{-4}(\partial_{\q\q} \zeta)^2$ ({\emph{i.e.}} variable shape along radial direction leads to negligible bending energy).     
\\

Choosing appropriate units of energy and length, one finds that the model energy (\ref{eq:FvKenergy}) depends on two dimensionless parameters (Eq.~6 of main text): 
\begin{equation}
\epsilon = \frac{BR^2}{YW^4} \sim \Big(\frac{\sqrt{tR}}{W}\Big)^4 \ \ ; \ \ \chi^{-1} = \frac{\Ksub R^2}{Y} \ , 
\end{equation}
whose physical meaning was discussed in the main text. As we explained there, the focus of our study is the asymptotic regime:
\begin{equation}
\epsilon^{-1} \gg \chi^{-1}  \gg 1  \ , 
\label{eq:inequality} 
\end{equation}
where the sheet is highly bendable and the substrate is barely deformable.  
%
A direct counting of dimensionless groups yields a third parameter, $W/R$, but in the F{\"o}ppl v{\'a}n K{\'a}rman (FvK) framework it can be absorbed into a global energy scale \cite{Hohlfeld15}.   

In order to find the displacement field it is to useful introduce the Euler-Lagrange equations of the energy functional (\ref{eq:FvKenergy}), which are essentially the FvK equations for the mechanical equilibrium of the sheet, subjected to the restoring force $-K(\zeta + r^2/2R)$: 
\begin{subequations}
\label{eq:div}
\begin{gather}
B \Delta^2 \zeta - \sigma_{rr}\partial_r^2\zeta - \tfrac2r \sigma_{r\q}\left(\partial_r -\tfrac1r\right) \partial_\q \zeta \nonumber \\
- \tfrac{1}{r^2} \sigma_{\q\q} \left(\partial_\q^2 \zeta + r\partial_r \zeta\right) = -K  [\zeta(r,\q) + r^2/2R] \ , \label{eq:normal} \\ 
\partial_r \sigma_{rr} + \tfrac1r \left(\partial_\q \sigma_{r\q}
+ \sigma_{rr} - \sigma_{\q\q}\right) =  0  \label{eq:divr}\ ,\\
\partial_r \sigma_{r\theta} + \tfrac1r \left(\partial_\q \sigma_{\q\q} + 2\sigma_{r\theta}\right) = 0  \ ,  \label{eq:divq} 
\end{gather}
\end{subequations}
where the Laplacian $\Delta \equiv \partial_r^2 + \frac1r\partial_r + \frac{1}{r^2}\partial_\q^2$. 
Eq.~(\ref{eq:normal}) is the $1^{st}$ FvK equation and expresses force balance in the normal direction; whereas Eqs.~(\ref{eq:divr},\ref{eq:divq}) form the $2^{nd}$ FvK equation, and express force balance on each infinitesimal piece of the film in the two directions locally tangent to the sheet \cite{LL86, timoshenko70, Mansfield}.

It is useful to consider the axially symmetric state, which is the solution for the case of an infinitely rigid substrate ($\chi^{-1} = \infty$). In this case the effect of the substrate can be expressed as a constraint -- the sheet must conform perfectly to a sphere. 
Clearly, the whole displacement field is axially symmetric, such that $\ru_\q=0$,
$\ru_r(r,\q) = \ru_r(r)$, and $\zeta(r,\theta) = \zeta_{sph}(r) \approx -r^2/2R$. Since 
any explicit dependence on $\q$ vanishes, the only non-vanishing components of the stress tensor remain $\sigma_{rr}(r)$ and $\sigma_{\q\q}(r)$. Furthermore, for an infinitely stiff substrate, the normal force balance (Eq.~\ref{eq:normal}) is replaced by the geometrical constraint, 
$\zeta(r,\theta) = -r^2/2R$, and the only remaining equation is the radial force balance, supplemented by the strain-displacement Eq.~(\ref{eq:strains}), akin to the compatibility condition \cite{Azadi12}. The radial displacement $\ru_r(r)$, as well as the radial and hoop stresses for the axisymmetric state are given in~\cite{Azadi12, Azadi14}. Crucially, the only stress scale that governs the magnitude of both stress components, $\srr$ and $\sqq$, of the axisymmetric state, is $Y (W/R)^2$, and the energy of such a state -- which is obviously dominated by strain -- is therefore $\sim YW^6/R^4$.  

\section{Simulation on sheets on a spherical Winkler foundation}
\subsection{Numerical model} We employ the ``bead-spring'' model of Seung and Nelson~\cite{NelsonSeung} implented in the MEMBRANE code developed by Skenpnek {\emph{et al.}}~\cite{Rastko}.  The sheet is modeled by an initially planar, equitriangular array of vertices $\xv_\alpha$, connected by nearest neighbor springs of rest length $a=1$ (all length scales are measured in units of the stress-free lattice spacing).  Upon deformation, in-plane strain energies are captured by spring stretching energy,
\begin{equation}
\label{eq: Estretch}
E_{\rm strain} = \frac{k}{2} \sum_{\langle \alpha \beta \rangle} \Big( |\xv_\alpha-\xv_\beta| - a\Big)^2 ,
\end{equation}
where the sum is taken over  nearest neighbor bonds and $k$ is a spring constant.  To model the bending energy of the sheet, normal vectors ${\bf n}_\mu$ are associated to each triangular plaquette, and the bending energy derives from the dihedral rotation of normals between adjacent plaquettes \begin{equation}
\label{eq: Ebend}
E_{\rm bend} = \frac{ \kappa}{2} \sum_{\langle\alpha\beta \rangle}  \big| {\bf n}_\alpha-{\bf n}_\beta\big|^2 ,
\end{equation}
where the sum is taken over neighbor (edge-sharing) plaquettes and $\kappa$ is an angular stiffness.  In the limit $a \ll W$, this discrete model tends towards the continuum limit of an elastic membrane with stretching and bending moduli, eq. \ref{eq:EnWindefine}, with modulii: 
\begin{equation}
Y = \frac{2 }{ \sqrt{3}} k; \  B = \frac{\sqrt{3}}{2} \kappa ,
\end{equation}
which we use to compare numerical simulations to analytical predictions. Finally, we consider an elastic interaction between the sheet and the spherical substrate of preferred radius $R$ of the form of a Winkler foundation that penalizes local deflection of the substrate in the radial (normal) direction, 
\begin{equation}
E_{\rm sub} = \frac{K}{2} \sum_{\alpha} \Big(|\xv_\alpha| - R \Big)^2
\end{equation}
where the origin is taken as the sphere center and $K$ is a stiffness parameter for the substrate.  In the limit of small triangles (and small strains) this energy tends towards the continuum form of eq.~\ref{eq:EnWindefine}, with
\begin{equation}
K_{\rm sub} = \frac{2}{ \sqrt{3} a^2} K .
\end{equation}
Given an initial configuration of the sheet, the simulation proceeds via energy minimization performed using the BFGS.  The convergence is determined by requiring that the sum over absolute value of derivative of energy on all vertices should be smaller than a threshold value.  A threshold level of $5 \times 10^{-7} ka$ was established  empirically by confirming that two benchmarks of the shape profile, the extent of the wrinkled zone and the wavelength of wrinkles, do not change by more than a few \% upon further decrease of the threshold.

Initial configurations are constructed by cutting planar triangular meshes from within a circular ``stencil'' of radius $W$.  To ensure that the discretization has minimal impact on simulated wrinkle patterns, we require sufficiently large ratios of $W/a$ so that the ultimate patterns satisfy the criteria $\lambda/a \gtrsim 10$ over the full range of simulated parameters.  Practically, this was achieved using sheets with $W= 300a$ corresponding to 326,467 vertices.  Energy relaxation begins from a state with the planar sheet projected orthographically onto the sphere of radius $R=3000a$.  Bending modulus and substrate stiffness were varied over a range of parameter values $\kappa/(ka^2) = 6\times 10^{-5} - 0.13$ (corresponding to $\epsilon=5\times 10^{-8} - 1.67 \times 10^{-4}$) and $K/k=5\times 10^{-6} -5\times 10^{-5}$ (corresponding to $\chi = 2.57 \times 10^{-3}-2.57 \times 10^{-2}$).  

\subsection{Simulation analysis}

Simulation results are analyzed from two morphological features: the radius $L$ of the unwrinkled 
core and the wrinkle wavelength $\lambda$, as well as the profiles of radial and hoop stress.  To analyze the shape, vertex positions are converted into the deflection profile:  $\Delta \zeta_\alpha =|\xv_\alpha| - R $, the normal deflection of a vertex from the sphere, its arc radius $r_\alpha = \arccos \big(\xv_\alpha \cdot \hat{z} /R \big)$ and azimuthal angle around the pole, $\theta_\alpha$.  The vertex positions are then binned into radial annuli of width $0.04W$, and then the mean number of wrinkles at a given arc radius is determined from counting the number of amplitude oscillations in the deflection profile plotted vs. $\theta$.  The length $L$ is determined via two algorithms.  In the first, which is implemented numerically for $L \lesssim 0.85W$, the radially averaged bending energy density of simulated sheets is plotted as a function of radial position.  The start of the wrinkled zone is taken as a radius at which the bending energy density has the largest second derivative (computed via interpolation), and the error bars as estimated as the full width, half maximum of this second derivative.  When the edge of the wrinkled zone approaches the edge of the sheet (i.e. when $L \approx W$), this numerical algorithm fails, and $L$ is determined from plots of $\Delta \zeta_\alpha$ vs. $r_\alpha$, as the extrapolation of the apparent wrinkle envelope at $r> L$ to $\Delta \zeta =0$.

The stress profiles in simulated sheets are computed using the virial formula for a vertex position $\alpha$
\begin{equation}
\sigma_{ij}(\alpha) = \frac{1}{2A_\alpha}\sum_{ \langle \alpha \beta \rangle } \left[ x_i(\alpha) - x_i(\beta) \right]f_j (\alpha, \beta)
\end{equation}
where the sum is taken over nearest (bonded) neighbors of $\alpha$, $A_\alpha$ is the area per vertex, and force from the $\alpha, \beta$ bond is  
\begin{equation}
{\bf f} (\alpha, \beta)= k \big(|{\bf x}_\alpha-{\bf x}_\beta| - a \big) \hat{x}_{\alpha \beta},
\end{equation}
where $ \hat{x}_{\alpha \beta} = ({\bf x}_\alpha-{\bf x}_\beta )/|{\bf x}_\alpha-{\bf x}_\beta|$. The radial and hoop components (plotted in Fig. 2 in the main text) are computed by projecting the stress tensor onto the local radial direction $\hat{r}_\alpha$ that points along the longitude from $r=0$ to $(r_\alpha, \theta_\alpha)$, where $\theta_\alpha = {\bf n}_\alpha \times \hat{r}_\alpha$.

\section{Basis of ITFT \label{sec:BasisITFT}} 

In main text we introduced the {\emph{inverted tension field theory}} (ITFT) for the spherical Winkler problem. In this exposition we explicitly assumed that, despite the strong oscillations (wrinkles) of the shape in the azimuthal direction, the residual strains and consequently stress field are governed by force balance equations that depend only on the radial direction, $r$. Here we justify this assumption by considering also the leading oscillatory terms in the displacement field and consequent strain components. We will show that the ITFT equations (11-13 in the main text) are essentially the leading order in an expansion of the FvK equations around an asymptotically strainless state, obtained in the singular limit (\ref{eq:inequality}), where the sheet becomes asymptotically strainless in the sense of the Gauss-Euler principle, Eq.~4 of main text.  

We further employ this asymptotic expansion to evaluate the total energy of the wrinkled state, 
$$U = \int_0^{2\pi}d\q\int_0^W rdr \ [u_{\rm bend} + \usub + u_{\rm strain}] \ , $$ in the limit (\ref{eq:inequality}). Normalizing energies by the factor $Y\tfrac{W^6}{R^4}$, which is the energy (up to a a numerical pre-factor) of the axisymmetrically deformed (unwrinkled, strained) state, we find that the leading contribution to the energy originates from bending the sheet and substrate deformation:       
\begin{equation}
U_{\rm bend} +U_{\rm sub} \to Y \frac{W^6}{R^4} \cdot \frac{\pi}{6}\sqrt{\eK} \ [1 + O(\sqrt{\eK})]  \ , 
\label{eq:limit-energy-o}
\end{equation}   
%
and the strain energy contributes only a sub-leading term:
\begin{gather}
U_{\rm strain}\to Y \frac{W^6}{R^4} \cdot \frac{2\pi}{3} \eK \ [-\log \eK + O(1)]  \  . 
\label{eq:limit-energy-sub-o}
\end{gather}   
The asymptotic expansion of the energy, given by Eqs.(\ref{eq:limit-energy-o},\ref{eq:limit-energy-sub-o}), provides a {\it self-consistent confirmation to the Gauss-Euler elastica principle} (Eq. 4 of the main text): 
The minimum energy of a GIC problem (in the limit~(\ref{eq:inequality}))  
is characterized by a dominant energy associated with  
bending and substrate deformation, and a sub-dominant energy due to residual strain.  
\\

We adopt a variational method to approximate the energy  
{\emph{via}} a parametric family of states that assumes radial wrinkles develop in an annular zone, $L<r<W$, anticipating that the sheet becomes fully covered with a diverging number of wrinkles ($L/W \to 0 \ ,\  m\to \infty$) in the singular limit (\ref{eq:inequality}). Hence, 
we have \beq       
L<r<W \ \ : \ \ \zeta(r,\theta) \approx \zeta_0(r) + f(r)\cos(m\theta) \ , 
\label{eq:wrinkle-form}
\eeq
where $\zeta_0(r) \approx \zeta_{sph}(r) \approx -r^2/2R$.
Our objective is to find the radial profile $f(r)$, the number of wrinkles $m$ (which may vary with $r$), the core radius $L$, and the other components of the displacement field, $\rmu_r$ and $\rmu_\q$, that minimize the total energy 
$U =\int_0^W rdr  \ [\ubend+\ustrain + \usub]$. 
Central to our reasoning is the high energetic price that comes with any level of strain due to the large contrast between the stretching and bending moduli. The overarching principle of our variational construction in this section is thus to keep strain at the {\it smallest possible level}, while requiring the sheet to remain as close as possible to the spherical substrate.  Implementation of this principle leads to a much stronger result, shown in the next subsection, that there exists a wrinkle pattern for which the {\it energy $U_{\rm strain}$ is negligible in the limit }(\ref{eq:inequality}) {\it in comparison with $U_{\rm bend}$ and $U_{\rm subs}$.} Finally, we will show that, despite this overall negligibility of strain, its minimization has a dominating effect on the separation between wrinkled and unwrinkled zones.  The spatial distribution of $u_{\rm strain}$ becomes concentrated in the limit (\ref{eq:inequality}) in the (internal) margin of the wrinkled zone, yielding a new type of ``stress focusing'' phenomenon, which underlies the core size, $L$.

\subsection{Asymptotically vanishing strain} 

\subsubsection*{Radial strain} 
We start by requiring the radial strain to vanish in the limit (\ref{eq:inequality}). 
Using Eqs.~(\ref{eq:strain-radial-1}, with $\Lambda=0$), the boundary conditions $\rmu_r(0) = 0$, and recalling our assumption that $L/W \to 0$, we obtain:
\begin{equation}
\varepsilon_{rr} \to 0 \ \  \Rightarrow \ \ \rmu_r \to -\tfrac{1}{6}\tfrac{r^3}{R^2}  \ \ 
\Rightarrow  \ \   \tDelta(r) \equiv -\tfrac{\rmu_r}{r} \to  \tfrac{1}{6}\tfrac{r^2}{R^2}
 \ , 
\label{eq:vanish-eps-rr}
\end{equation}  
as was mentioned already in the main text. 

The above limit characterizes the leading term in the radial displacement, which remains finite in the limit (\ref{eq:inequality}). Establishing the hierarchical structure of the energy, as described in the next subsection, requires consideration of the displacement beyond leading order. Anticipating this necessity, we generalize the above expression for $\tDelta(r)$ as 
\begin{equation}
\tDelta(r) \to   \frac{1}{6}\frac{r^2}{R^2} + g_\Delta(r) 
\ , 
\label{eq:vanish-eps-rr-n}
\end{equation}   
where the function $g_\Delta(r)$ is required to vanish in the limit (\ref{eq:inequality}).  

\subsubsection*{Azimuthal strain} Inserting the ansatz (\ref{eq:wrinkle-form}), into Eq.~(\ref{eq:hoopstrain}), and requiring hoop strain to vanish we obtain from the non-oscillatory ({\emph{i.e.}} $\theta$-independent) part of $\varepsilon_{\q\q}$:  
\begin{equation}
\varepsilon_{\q\q} \to 0 \ \  \Rightarrow \ \ \tfrac{1}{4r^2}f^2m^2 \to \tDelta(r)  \ , 
\label{eq:vanish-eps-tt}
\end{equation}   
where the azimuthal confinement $\tDelta(r)$ is given by Eq.~(\ref{eq:vanish-eps-rr}). Anticipating again the analysis in the next subsection, we complement the above expression with a higher-order term: 
\begin{equation}
\tfrac{1}{4r^2}f^2m^2 \to \tDelta(r) + g_f (r) \ , 
\label{eq:vanish-eps-tt-n}
\end{equation}  
where the function $g_f(r)$ is required to vanish in the limit (\ref{eq:inequality}).   

At a leading order approximation, where $\tDelta(r) = \tfrac{1}{6}\tfrac{r^2}{R^2}$, Eq.~(\ref{eq:vanish-eps-tt-n}) is a {\emph{``slaving condition''}} between the amplitude 
$(f)$ and wavelength ($\lambda = 2\pi r/m$) of wrinkles \cite{Davidovitch11,Hohlfeld15}: both $f$ and $\lambda$   
become indefinitely small in the limit (\ref{eq:inequality}), but their ratio must remain finite in order to allow collapse of hoop and radial components of the strain tensor.

The oscillatory ({\emph{i.e.}} $\theta$-dependent) part of Eq.~(\ref{eq:hoopstrain}) shows that an asymptotically vanishing hoop strain implies an additional constraint on the  
azimuthal displacement $\rmu_\q$:
\begin{equation}
\varepsilon_{\q\q} \to 0 \ \  \Rightarrow \ \ \rmu_{\q} \to + \frac{r}{2m}\tDelta(r) \sin(2m\q) \ , 
\label{eq:vanish-eps-tt-oscil}
\end{equation}    
where we used Eq.~(\ref{eq:vanish-eps-tt-n}). Once again, we allow for higher-order contributions to $\rmu_\q$ as we elaborate in the following subsection: 
\begin{equation}
\rmu_{\q}  \ \to \  r g_{\q 1}(r) \sin(m\q)  \ + \  \frac{r}{2m}[\tDelta(r)  + g_{\q 2} (r) ]    \sin(2m\q)  
 \ , 
\label{eq:vanish-eps-tt-oscil-n}
\end{equation}    
where the functions $g_{\q 1}(r),g_{\q 2}(r)$, are required to vanish (faster than $m^{-1}$) in the limit (\ref{eq:inequality}).   
\\

\subsubsection*{Shear strain} 
The last component of the strain tensor is the shear. Inserting the ansatz (\ref{eq:wrinkle-form}), into Eq.~(\ref{eq:strain-shear-1}), and using the above result, Eq.~(\ref{eq:vanish-eps-tt}), one may be tempted to conclude that the shear component approaches a finite value: $\varepsilon_{r\q} \approx 
\tfrac{1}{2r} \partial_r\zeta\partial_\q \zeta \approx \sqrt{\tDelta(r)}(\tfrac{r}{R})\sin(m\q)$. However, an asymptotically vanishing strain is retained (at an ultimately nominal energetic cost) through an oscillatory addition to the radial displacement:
 \begin{gather}
\varepsilon_{r\q} \to 0  \Rightarrow   \ \  
\rmu_r  \to  - r\tDelta(r)  \ \  \nonumber \\
 \ \  \ \ \ \ \ \ \ \ \ \ \ \ \ \ \  \ + \
\frac{1}{m} \frac{2r^2}{R} \ [ \sqrt{\tDelta(r)} + g_{r1}(r)] \ \cos(m\q)  
  \ ,  
\label{eq:vanish-eps-rt}
\end{gather}   
where $g_{r1}(r)$ is yet another function which is required to vanish in the limit (\ref{eq:inequality}).  
Note that the oscillatory correction to the radial strain, $\varepsilon_{rr}$, induced by the necessity to eliminate shear, scales inversely with $m$, and is thus sufficiently small such that the radial strain still vanishes in the limit (\ref{eq:inequality}). As we will show in the next subsection, this oscillatory contribution to $\rmu_r$ underlies the ``geometric stiffness'', $K_{\rm curv}$, Eq.~(14) of the main text
\cite{Paulsen16,Hohlfeld15}.   

Reconsidering Eq.~(\ref{eq:strain-shear-1}), one may jump to the conclusion that the residual shear strain is  
$\tfrac{1}{2}\partial_r\rmu_\q = \tfrac{1}{8m}\tfrac{r^2}{R^2}\sin(2m\q) \sim O(m^{-1})$. 
Since $m\to\infty$ in the limit (\ref{eq:inequality})), such a contribution seems harmless; however, we will see later that it is inconsistent with an asymptotic vanishing of other strain components. Inspection of the various terms in (\ref{eq:strain-shear-1}) reveals that the actual shear strain can be reduced to terms of order $O(m^{-2})$ or higher, by including a higher-order contribution to the $\rmu_r$:   
\begin{gather} 
\rmu_r  \to  - r\tDelta(r)   +  \frac{1}{m} \frac{2r^2}{R} \ [\sqrt{\tDelta(r)} + g_{r1}(r)] \cos(m\q)    \nonumber \\ +  
\frac{1}{m^2} r\Big[    -\frac{3}{4}\tDelta(r) -\frac{1}{4} r  \frac{d}{dr}  \tDelta(r) - g_{f}(r)  +  g_{r2}(r)   ] \cos(2m\q)   \ , 
\label{eq:vanish-eps-rt-h} 
\end{gather} 
where we introduced a final function, $g_{r2}(r)$, that must also vanish in the limit (\ref{eq:inequality}).

\subsection{Dominant and sub-dominant energies}
\label{domandsub}
We showed above that requiring all strain components to asymptotically vanish provides strict constraints on the form of the displacement field $[\rmu_r,\rmu_\q,\zeta]$, as expressed in Eqs.~(\ref{eq:vanish-eps-rr-n}, \ref{eq:vanish-eps-tt-n},\ref{eq:vanish-eps-tt-oscil-n},\ref{eq:vanish-eps-rt-h}). The ability to eliminate strain suggests that in the limit (\ref{eq:inequality}), the total energy is not  ``equipartitioned'' between the various deformation modes, but instead -- similarly to buckling and wrinkling phenomena in one-dimensional (1D) systems -- becomes dominated by the bending and substrate deformation costs, $u_{\rm bend}$ and $\usub$, rather than by $u_{\rm strain}$. We will prove this nontrivial claim in a self-consistent manner, evaluating first the limit value of $\int_0^{2\pi}d\q\int_0^W rdr  \ (u_{\rm bend} + \usub)$, then the residual strain energy $\int_0^{2\pi}d\q\int_0^W rdr \ u_{\rm strain}$, and finally show that the energetic cost of the latter is negligible in comparison to the former. More precisely, expanding the total energy, $U$, in the {\emph{two}} small parameters, $\eK$ and $\eoK$, we will show that: 
\begin{equation}
\frac{U}{YW^6/R^4} = C_0 \eK^{1/2}[1 \ + \ C_1  \eoK^{1/2}]
 + C_2 {\eK}  
\   + \cdots   \ ,  
\label{eq:expansion-energy}
\end{equation}
where the constant $C_0$ is obtained by minimizing $u_{\rm bend}$ and $\usub$ alone. Minimization of the energy stored in the residual strain,  
which ultimately determine the radius $L$ of the unwrinkled core, will be carried out in the next subsection, yielding the constants $C_2$ and $C_1$. The latter will be shown to acquire a logarithmic correction, which is intimately related to the core size $L$. 
\\   

Assuming a total energy $U$ dominated by $\ubend$ and $\usub$, the mechanics of a single hoop becomes analogous to an elastic ring of radius $r$ attached to a substrate of stiffness $\Ksub$, which is subjected to confinement $\tDelta(r)$ ({\emph{i.e.}} radial contraction, $r \to r(1-\tDelta(r))$. Hence, hoops become wrinkled, and their response 
is described by Eq.~(11) of the main text. Namely, 
each latitude, labeled by $0<r<W$, responds as a 1D {\emph{elastica}} confined by 
$\tDelta = \tfrac{1}{6}\tfrac{r^2}{W^2}$, with wrinkles of wavelength $\lambda = 2\pi (\tfrac{B}{\Ksub})^{1/4}$ (note that for low deformability ($\chi^{-1}\gg 1$), we have  $\Keff \approx \Ksub$).  Anticipating that the wrinkle pattern asymptotically covers the whole sheet, 
we find that the asymptotic value of the energy in the limit (\ref{eq:inequality}) is: 
\begin{gather}
U \to \int_0^{2\pi} d\q \int_0^W rdr \ 2\sqrt{B\Ksub} \ \tDelta(r) = \frac{\pi W^6}{6R^4}\sqrt{B\Ksub}   \nonumber 
\\  \Rightarrow \ \ C_0 = \frac{\pi}{6}  ,  
\label{eq:limit-energy}
\end{gather}   
Normalizing this result by $Y W^6/R^4$ we find the leading term in eq. (\ref{eq:limit-energy-o}).  Evaluating the next terms in the expansion (\ref{eq:expansion-energy}) requires consideration of small, finite values of $\eK$ and $\eoK$, for which a core zone of radius $L$ remains unwrinkled, and we thus consider three corrections to the leading energy (\ref{eq:limit-energy}): 

\noindent (i) a (positive) contribution from the residual strain in the wrinkled zone: $U_{\rm strain} = \int_0^{2\pi} d\q\int_L^W rdr \ u_{\rm strain}$; 

\noindent (ii) a (negative) contribution to the leading energy: $-  \int_0^{2\pi} d\q\int_0^L rdr \ (u_{\rm bend} + \usub)$. 

\noindent (iii) a (positive) contribution, dominated by strain, from the energy of the unwrinkled core: 
$U_{\rm core} = \int_0^{2\pi} d\q\int_0^L rdr \ u_{\rm strain}$.

As we show in the next subsection, the balance between these next-order terms (which turns out to be dominated by (i) and (iii)) yields the scaling law 
for $L$ (Eq.~(8) of the main text), and establishes the energetic sub-dominance of the strain in comparison to bending and substrate deformation.


\subsection{Hierarchy of FvK equations} 
In contrast to a general solution of FvK equations 
(\ref{eq:div}), which requires the equations to be solved {\emph{simultaneously}}, the theory presented here has a {\emph{sequential}} character, reflecting the anticipated separation of energy scales and the consequent hierarchical structure inherited to the corresponding Euler-Lagrange ({\emph{i.e.}} FvK) equations. Hence, one must choose carefully the order at which to address the three force balance equations, and the corresponding compatibility equations. In the analysis below, steps 0-2 repeat the arguments already mentioned in the main text, and steps 2-7 prove the consistency of neglecting various $\theta$-dependent terms from the strain and force balance equations.  
\\


\noindent {\bf 0. Shape and wavelength constraints:} Anticipating  the energy to be dominated by bending and substrate deformation, we require a solution of FvK equations of the form (\ref{eq:wrinkle-form}), where $\zeta_{0}(r) \to \zeta_{sph}(r)\approx -r^2/2R$, as implied by the high energetic cost of substrate deformations, and a wrinkle wavelength $\lambda = 2\pi(\frac{B}{\Ksub})^{1/4}$, which derives from the  balance of these two highest energetic costs. In other words, a wrinkled shape with  
$\zeta_0(r) \neq \zeta_{sph}(r)$ or $\lambda \neq 2\pi(\frac{B}{\Ksub})^{1/4}$, has an energy that is larger than the leading energy,   
$C_0\sqrt{\eK}$ (Eq.~\ref{eq:expansion-energy}) in the limit (\ref{eq:inequality}),  
and is therefore excluded. Hence, the starting point for our analysis of the FvK Equations is the shape (\ref{eq:wrinkle-form}), with $m(r)$ and $\zeta_0(r)$: 
\begin{equation}
\zeta_{0}(r) \to \zeta_{sph}(r)\approx -r^2/2R \ \ \ \  \ m(r) =  \eoK^{-1/4}\big( \frac{r}{W} \big)   \ .  
\label{eq:m-scale-n}
\end{equation}
We will comment later on the unavoidable energetic cost associated with the variation of the wrinkle number, $m(r)$, between integer values.
Clearly, we expect also the wrinkle amplitude $f \to 0$, otherwise azimuthal derivative such as $\partial_\q\zeta$ will diverge, leading to unphysical divergence of shear and hoop strains. 

In the forthcoming analysis, we will find it more convenient to carry out the expansion in $\eK$ and $m^{-1}$; Eq.~(\ref{eq:m-scale-n}) is then used to transform the final expression into an expansion in $\eK$ and $\eoK$.       
\\

\noindent {\bf 1. Normal force balance:} Normal force balance, Eq.~(\ref{eq:normal}), expresses the balance between bending and substrate deformations, and hence, it is the first one to analyze in the hierarchy. As was already mentioned, 
substituting the ansatz (\ref{eq:wrinkle-form}) with (\ref{eq:m-scale-n}), Eq.~(\ref{eq:normal}) reduces to the force balance on a hoop that becomes wrinkled under radial confinement:
\begin{gather}
B \frac{m^4}{r^4} f  + \sigma_{\q\q} \frac{m^2}{r^2} f  + \Ksub f = 0  \ , 
\label{eq:normal-new} 
\end{gather}
where we used the fact that in the limit (\ref{eq:inequality}) the radial shape $\zeta_0(r) \to -r^2/2R$, and each of the forces, $B \nabla^2 Tr (\kappa_{ij})$, and $-\sigma_{ij}\kappa_{ij}$,  
is dominated by the component $\kappa_{\q\q} \sim - r^{-2} m^2 f\cos(m\q)$ of the curvature tensor. The neglected terms contribute at higher order in $m^{-1}$ and $\eK$ \cite{Hohlfeld15}, and their primary effect is to ``renormalize'' the substrate stiffness: $\Ksub \to \Keff$ (Eq.~14 of the main text, see step 5 below).  
Eq.~(\ref{eq:normal-new}) has multiple solutions, however, the one that minimizes the energy of hoops is obtained by balancing the two resistive forces (bending and substrate).  
Hence, 
we obtain an equation for the hoop stress: 
\begin{gather}   
[1^{st} \ {\rm FvK}]  \ + \ [\lambda = 2\pi \Big(\frac{B }{\Keff}\Big)^{1/4}] \nonumber  \\ 
\Rightarrow  \ \ \ \sigma_{\q\q} \to -2\sqrt{B\Keff} \approx -2Y\Big(\frac{W}{R}\Big)^2\sqrt{\eK} \ .  
\label{eq:FvK1-nn}
\end{gather}  
\\


\noindent {\bf 2. Radial force balance:} Turning now to the next-order contribution to the energy, we consider the in-plane, radial force balance eq. (\ref{eq:divr}). Addressing first the non-oscillatory ($\theta$-independent) terms, we obtain: 
\begin{equation}
\partial_r(r\sigma_{rr}) - \sigma_{\q\q} = 0 \ . 
\label{eq:FvK2-nn1}
\end{equation} 
In standard tension field theory, the residual compressive stress is neglected, and the magnitude of $\sigma_{rr}$ is determined by the tensile boundary load. However, in our problem, where the load-free boundary conditions imply $\sigma_{rr}(W)=0$, we must retain the residual compression, $\sigma_{\q\q}$, which is not relieved by wrinkles, in the in-plane force balance.
In other words, under load-free conditions, 
the term $\sigma_{\q\q} = -2Y(\frac{W^2}{R^2})\sqrt{\eK}$ becomes the leading non-homogenous source in the FvK equations (\ref{eq:div}), and thus sets the magnitude of the other stress components, $\sigma_{rr}$ and  $\sigma_{r\theta}$.   

Using Eq.~(\ref{eq:FvK1-nn}) for $\sigma_{\q\q}$, the boundary condition $\sigma_{rr}(W)=0$, the strain-displacement compatibility Eq.(\ref{eq:strain-radial-1}) with $\zeta(r) \approx \zeta_{sph} \approx -r^2/2R$, and expanding Eq.~(\ref{eq:FvK2-nn1}) in $\eK$, we obtain Eqs.~(\ref{eq:vanish-eps-rr},\ref{eq:vanish-eps-rr-n}) for $\tDelta(r)$, with: 
\begin{equation}
g_\Delta(r) = 2 \sqrt{\eK} \Big(\frac{W}{R} \Big)^2 \Big[1 + \frac{W}{r} \log \Big(\frac{L_0}{r}\Big) \Big]     \ . 
\label{eq:gDelta}
\end{equation}  
This gives rise to the leading contribution of the radial stress: 
\begin{equation}
\sigma_{rr} \to 2Y \Big(\frac{W}{R} \Big)^2 \sqrt{\eK} \Big(\frac{W}{r} - 1\Big) \ . 
\label{eq:lead-radial-stress}
\end{equation}  
Notice that we cannot integrate Eq.~(\ref{eq:FvK2-nn1}) down to $r=0$ (where one must have $\rmu_r(0)=0$) due to the $1/r$ divergence in Eq.~(\ref{eq:gDelta}). This divergence is the mechanism that determines the radius $L$ of an unwrinkled core. As we will see below, matching analysis with the strained, unwrinkled core, relates the remaining constant, $L_0$, on the RHS of Eq.~(\ref{eq:gDelta}) with the core radius $L$. 
\\

Before turning to the last force balance equation (\ref{eq:divq}) 
we must first address some strain-displacement compatibility conditions, in the first two FvK equations (\ref{eq:normal},\ref{eq:divr}), that were neglected in steps 0-2 (and in the main text).  
\\

\noindent {\bf 3. Compatibility of hoop strain:} Compatibility of Eq.~(\ref{eq:FvK1-nn}) with stress-strain (\ref{eq:stresses}) and strain-displacement (\ref{eq:hoopstrain}), can be Fourier-decomposed into two equations. 
The non-oscillatory ($\theta$-independent) part yields the function $g_f(r)$: 
\begin{equation}
g_f(r) = -2 \Big(\frac{W}{R} \Big)^2 \sqrt{\eK}  \  . 
\label{eq:gf}
\end{equation}
Equations~(\ref{eq:vanish-eps-tt-n},\ref{eq:gf}) consists a ``relaxed'' slaving condition, in contrast with its counterpart in standard tension field theory, where one assumes 
a perfect ``inextensibility'' along latitudes ({\emph{i.e.}} neglecting any deviations between the excess length of wrinkles, $\tfrac{1}{4r^2} m^2f^2$, and the hoop confinement $\rmu_r/r$) \cite{King12, Hohlfeld15}. Taken together, Eqs.~(\ref{eq:vanish-eps-rr-n},\ref{eq:gDelta}) and (\ref{eq:vanish-eps-tt-n},\ref{eq:gf}) are a central outcome of our ITFT analysis; we will show below that the two functions $g_{\Delta},g_f$, determine the leading terms in the residual strain components, 
$\varepsilon_{rr},\varepsilon_{\q\q}$, which underlie the next order terms, $C_1\eK$ and $C_2 \eoK$, in the energy expansion (\ref{eq:expansion-energy}). 

 

Turning now to the compatibility of the oscillatory part of the hoop strain ~(\ref{eq:hoopstrain}), we note that an $O(\sqrt{\eK})$ oscillatory contribution to the hoop strain can be avoided if the azimuthal displacement $\rmu_\q$ (\ref{eq:vanish-eps-tt-oscil-n}), is characterized by a function $g_{\q 2}(r)  = g_f(r)$, up to corrections that 
vanish faster than $\sqrt{\eK}$ in the limit (\ref{eq:inequality}). The origin of the additional function $g_{\q 1}(r)$, and the higher order corrections to $g_{\q 2}(r)$, will be addressed in the next paragraphs. . 
\\


\noindent {\bf 4. Shear:} In our analysis of Eqs.~(\ref{eq:FvK1-nn}) and (\ref{eq:FvK2-nn1}), we simplified the original FvK equations (\ref{eq:normal},\ref{eq:divr}) by ignoring the forces associated with the shear stress. This approach seems to be particularly problematic for the radial force balance, where the effect of the azimuthal derivation, $|\partial_\q (\cdot)| \approx m|(\cdot)|$, 
can make the shear-induced radial force 
$\tfrac{1}{r}\partial_\q\sigma_{r\q}$ larger than the two other forces, thus invalidating the dominant balance underlying Eq.~(\ref{eq:FvK2-nn1},\ref{eq:gDelta}). 
%
To avoid such a scenario, the magnitude of the shear stress in the limit (\ref{eq:inequality}) must not exceed 
$O(m^{-2})$; 
a shear that scales as $m^{-1}$ will give rise to $O(1)$ radial force that can be balanced only by an $O(1)$ contribution to $\sigma_{rr}$ or $\sigma_{\q\q}$, and is therefore inconsistent with an asymptotically vanishing strain. This reasoning underlies 
the inclusion of a second oscillatory term ($\propto \cos(2m\q)$) in the radial displacement, Eq.~(\ref{eq:vanish-eps-rt-h}). 
\\

\noindent {\bf 5. Oscillatory radial stress and curvature-induced stiffness:} 
Inspection of the radial displacement~(\ref{eq:vanish-eps-rt-h}), together with eq.~(\ref{eq:vanish-eps-tt-n}) yields
\begin{gather}
\sigma_{rr} =  Y\frac{W^2}{R^2}  \Big\{  \ 2 \sqrt{\eK} \Big(\frac{W}{r} - 1\Big)  \  \nonumber \\
 +  \ 
\frac{2}{m} \frac{R}{W^2} r \sqrt{\tDelta(r)} \cos(m\q)   \  +  
  \cdots   \ \Big\} \ , 
\label{eq:stress-asy-2}
\end{gather} 
where ``$\cdots$'' stands for contributions that are $O\eK$ or $O(m^{-2})$ in the limit (\ref{eq:inequality}).
Using $\sqrt{\tDelta(r)} \to fm/2r$ (eq.~\ref{eq:vanish-eps-tt-n}) and substituting the oscillatory part of $\sigma_{rr} $ ($\propto \ \cos (m \theta)$) into the first FvK equation, leads to 
another term in the LHS of 
eq. (\ref{eq:normal-new}), $Y\tfrac{d^2\zeta_{sph}}{dr^2} f(r)$. This term underlies a renormalization of the stiffness $\Ksub \to \Keff = \Ksub+\Kcurv$ (Eq.~14 of the main text).

Reconsidering the oscillatory terms in the radial force balance (\ref{eq:divr}), Eq.~(\ref{eq:stress-asy-2}) enables us to evaluate the leading contribution to the shear stress. As we shall see below (step 6), the oscillatory contribution to the hoop stress must vanish faster than $O(m^{-1})$, and hence cannot balance the $O(m^{-1})$ associated from the oscillatory part of $\tfrac{\partial}{\partial r} (r \sigma_{rr})$ in the radial force balance equation. Thus, we obtain: 
%
%
\begin{equation}
\sigma_{r\q} = Y \frac{W^2}{R^2}  \Big\{ -\frac{2}{m^2} \frac{R}{W^2} \frac{d}{dr}[r^2\sqrt{\tDelta(r)}] \sin(m\q) + \cdots \Big\} \ . 
\label{eq:stress-asy-3}
\end{equation}  
Inspection of the relevant strain-displacement relation (\ref{eq:strain-shear-1}) shows that eq.~(\ref{eq:stress-asy-3}) can be satisfied by multiple choices of pairs $g_{\q 1}(r),g_{r1}(r)$ in the azimuthal and radial displacements, (\ref{eq:vanish-eps-tt-oscil-n},\ref{eq:vanish-eps-rt-h}), respectively. This degeneracy will be resolved next, once we address the azimuthal force balance equation.   
\\



\noindent {\bf 6. Azimuthal force balance:} Resolving the form and scale of the shear stress puts us in a position to address the last force balance equation (\ref{eq:divq}). This equation reflects a balance between the force $\tfrac{1}{r}\partial_\q\sigma_{\q\q}$ and forces derived from the shear stress, $(\tfrac{2}{r}+ \partial_r)\sigma_{r\q}$. However, our discussion in step 4 above implies that the shear-induced azimuthal force must be $O(m^{-2})$; hence, since $|\partial_\q (\cdot)| \approx m|(\cdot)|$, we conclude that the magnitude of the oscillatory part of the hoop stress must be $O(m^{-3})$ in the limit (\ref{eq:inequality}).  

Inspecting the strain-displacement relation (\ref{eq:hoopstrain}), and the asymptotic forms of the azimuthal and radial displacements, Eqs.~(\ref{eq:vanish-eps-tt-oscil-n},\ref{eq:vanish-eps-rt-h}), we find that cancellation of $O(m^{-2})$ contributions to $\sigma_{\q\q}$ 
yields the leading terms in the expansions of $g_{\q 1}(r)$ and $g_{\q 2}(r)$:  
\begin{gather}
g_{\q 1}(r) = -\frac{2}{m^2} \frac{r}{R}\sqrt{\tDelta(r)} \label{eq:gq1}   \\
g_{\q 2}(r) =- 2 \Big(\frac{W}{R} \Big)^2 \sqrt{\eK} + \frac{1}{4m^2} \big[3 \tDelta(r) +  r  \frac{d}{dr}  \tDelta(r) \big]   \label{eq:gq2}  \ .   
\end{gather}  
Equation~(\ref{eq:gq1}) lifts the degeneracy in the form of the displacement field that was leftover from the above calculation of the leading order of $\sigma_{r\q}$. Employing Eq.~(\ref{eq:stress-asy-3}), the strain-displacement compatibility (\ref{eq:strain-shear-1}), the asymptotic forms of the displacements (\ref{eq:vanish-eps-tt-oscil-n},\ref{eq:vanish-eps-rt-h}), and Eq.~(\ref{eq:gq1}), we find: 
\begin{equation}
g_{r 1}(r) = \frac{1}{m^2} \frac{1}{r} \frac{d}{dr}[r^2\sqrt{\tDelta(r)}] \label{eq:gr1}   \ . 
\end{equation}       
\\

Finally, Eq.~(\ref{eq:FvK1-nn}), together with Eqs.~(\ref{eq:divq},\ref{eq:stress-asy-3}) provide us with the leading behavior of the non-oscillatory and oscillatory parts of the hoop stress in the limit (\ref{eq:inequality}): 
\begin{gather}
\sigma_{\q\q} =  Y\frac{W^2}{R^2} \Big\{  \ -2 \sqrt{\eK}  \   \nonumber \\ -   
\frac{2}{m^3} \frac{R}{W^2} \Big( (2 + r\frac{d}{dr}) \frac{d}{dr} \big[r^2\sqrt{\tDelta(r)} \big] \Big) \cos(m\q)   \  +  
  \cdots   \ \Big\} \ . 
\label{eq:stress-asy-4}
\end{gather}  
Attempting to determine the terms in the displacement field that give rise to the $O(m^{-3})$ oscillatory contribution to the hoop stress, we encounter once again the inherent hierarchical nature of our analysis. Similarly to our experience with the $O(m^{-2})$ shear stress (Eq.~\ref{eq:stress-asy-4}), there is a multitude of possible high-order combinations in the asymptotic forms of the radial and azimuthal displacements, Eqs.~(\ref{eq:vanish-eps-tt-oscil-n},\ref{eq:vanish-eps-rt-h}) that are compatible with Eq.~(\ref{eq:stress-asy-4}). In order to remove this degeneracy, one must proceed to analyze higher-order corrections to the force balance equations.     
\\

\noindent {\bf 7. Higher order contributions:}  
The asymptotic form of the displacement field, Eqs.~(\ref{eq:wrinkle-form},\ref{eq:vanish-eps-rr-n},\ref{eq:vanish-eps-tt-n},\ref{eq:vanish-eps-tt-oscil-n},\ref{eq:vanish-eps-rt-h}), and Eqs.~(\ref{eq:m-scale-n},\ref{eq:gDelta},\ref{eq:gf},\ref{eq:gq1},\ref{eq:gq2},\ref{eq:gr1}), provides 
the leading 
terms of the stress tensor, Eqs.~(\ref{eq:stress-asy-2},\ref{eq:stress-asy-3},\ref{eq:stress-asy-4}). With these stress components, all FvK force balance equations are satisfied up to forces of $O(m^{-2})$ or higher, in a manner that is compatible with the strain-displacement relations. As we will show below, the strain energy stored in this stress field gives rise to the coefficients $C_1,C_2$ in the energy expansion (\ref{eq:expansion-energy}), and this level of the expansion is thus sufficient for the purpose of this paper. 


Computing higher-order contributions to the stresses is possible but becomes a tedious task.  
In order to get an idea on the level of complexity, consider the normal forces balance, Eq.~(\ref{eq:normal}), which at $O(m^{-2})$ involves the $O(m^{-3})$ oscillatory part in $\sigma_{\q\q}$ that multiplies $-\tfrac{1}{r^2}m^2f\cos(m\q)$, as well as the product $-\sigma_{r\q}  \tfrac{d\zeta_{sph}}{dr} \tfrac{1}{r} m f\sin(m\q) $. Other contributors to the normal force at $O(m^{-2})$ come from the (mixed derivatives part of) bending force, $m^{-2} \tfrac{1}{r^2}\tfrac{d^2f}{dr^2}$, and a  contribution to the product $\sigma_{rr}\tfrac{d^2}{dr^2}$ from the $O(m^{-2})$ correction to $\sigma_{rr}$ that was left out in Eq.~(\ref{eq:stress-asy-2}). The resulting equation has to be solved simultaneously with the $O(m^{-2})$ solutions of radial force balance (which involves $O(m^{-2})$ terms in $\sigma_{rr}$ and 
$O(m^{-3})$ terms in $\sigma_{r\q}$), and the $O(m^{-2})$ solution of the azimuthal force balance (\ref{eq:divq}), which yielded Eq.~(\ref{eq:stress-asy-4}). 
%
These force balance equations must be solved simultaneously with the corresponding strain-displacement compatibility equations, and one has to find higher order corrections to the various $g(r)$ functions, and introduce necessary new terms into the displacement field ({\emph{i.e.}} a $\propto \cos(m\q)$ in $u_\q$ whose coefficient must be $O(m^{-4})$ or higher). Pursuing the expansion at higher orders, one will likely find it necessary to complement the displacement field with terms that oscillate at higher frequencies, namely $\cos(nm\q)$ and $\sin(nm\q)$ with $n>2$. 
\\

\subsection{The residual strain energy}
Recalling that our analysis consists of expansion in the two dimensionless parameters, $\eK$ and $\eoK$, Eqs.~(\ref{eq:stress-asy-2},\ref{eq:stress-asy-3},\ref{eq:stress-asy-4}) for the stress components can be written formally as products of asymptotic series in these two parameters. For instance, the expansion of the radial stress is: 
\begin{multline}
\sigma_{rr} =  Y \frac{W^2}{R^2}  \bigg\{  \Big[2 \sqrt{\eK} \Big(\frac{W}{r} - 1\Big)  + o(\sqrt{\eK})\Big] \times   \big[1 + o(m^{0}) \big]
    \\ +  \big[1 + o(\eK^0) \big] \times  \big[\frac{2}{m} \frac{R}{W^2} r \sqrt{\tDelta(r)} \cos(m\q) + o(m^{-1})\big]   \bigg\} , 
\label{eq:stress-asy-12}
\end{multline} 
where we used the common notations of asymptotic analysis: $$X = o(\cdot) \Rightarrow X \ll (\cdot) \ \ ;  \ \ X = O(\cdot) \Rightarrow X \sim (\cdot) \ ,  $$
and analogous structures for $\sigma_{r\q}$ and $\sigma_{\q\q}$. Inspection of these expansions reveals that upon integrating $U_{\rm strain} = \int_0^{2\pi}\int_0^W r drd\q \ u_{\rm strain} $, with $u_{strain}$ given by Eq.~(\ref{eq:EnWindefine}), the dominant terms in the double expansion originate from the non-oscillatory parts in $\sigma_{rr}$ and $\sigma_{\q\q}$, and the oscillatory part of $\sigma_{rr}$. For the last two terms, the spatial integration can be easily calculated; However, the first term gives rise to a logarithmically-diverging integral, which requires the introduction of a small cut-off scale $L >0$: 
\begin{equation}
U_{\rm strain} =  \pi Y\frac{W^6}{R^4}  \Big\{ \ 4\ \eK\cdot  \Big(-1 +  \log\frac{W}{L}\Big)   +   \frac{1}{12} \sqrt{\eoK} \   + \cdots \ \Big\} \ , 
\label{eq:Ustrain-1}  
\end{equation} 
where ``$\cdots$'' refers to higher-order contributions in $\eK$ or $\eoK$. The above expression for $U_{\rm strain}$, requires us 
to find the cut-off scale $L$, in order to establish the asymptotic energy expansion (\ref{eq:expansion-energy}).

As we anticipated, the logarithmic divergence in eq.~(\ref{eq:Ustrain-1}) indicates the presence of an unwrinkled zone of radius $L$, which vanishes in the limit (\ref{eq:inequality}), within which the sheet retains a small level of strain. In order to find $L$, it is useful to treat it as an independent variable, consider the total asymptotic energy $U[\eK,\eoK;L]$, assuming $L\to 0$ in the limit (\ref{eq:inequality}), and find $L$ by minimizing $U$. Following this procedure, and recalling that  we are only interested here in a leading order analysis of the limit (\ref{eq:inequality}),  we identify three terms that need to be addressed: 
\\

\noindent {\emph{(a)}} The {\emph{singular}} term in Eq.~(\ref{eq:Ustrain-1}), which is 
\begin{equation} 
U_{\rm sing} = 4\pi Y\frac{W^6}{R^4} \ \eK \log\Big(\frac{W}{L} \Big) \  . 
\label{eq:USing}
\end{equation}
Although this term scales only as  $\eK$, which is much smaller than the leading $\sqrt{\eK}$ contribution from bending and substrate deformation, the singular dependence on $L$ forces us to include it in the analysis. 
\\

\noindent {\emph{(b)}} The regular parts of the energy, which can be expanded as a Taylor series in $\eK,\eoK$, and $L$, include the leading order term (\ref{eq:limit-energy}), and all terms in $U_{\rm strain}$ except 
$U_{\rm sing}$.  This can be written as: 
\begin{multline} 
U_{\rm reg} =  U_{\rm reg}^0   + \pi Y\frac{W^6}{R^4} \ \bigg\{ - \sqrt{\eK} \ \frac{1}{6}\Big(\frac{L}{W}\Big)^3  \\  -  \eK  \ 4\Big(\frac{L}{W}\Big)^2 - \sqrt{\eoK} \  \frac{1}{12}  \Big(\frac{L}{W}\Big)^3 \bigg\} \ , 
\label{eq:UReg}
\end{multline}
where $U_{\rm reg}^0$ depends on $\eK$ and $\eoK$, but not on $L$.  
\\

\noindent {\emph{(c)}} The energy of a strained, unwrinkled core of radius $L$. Since force balance at the border $r=L$ between the wrinkled and unwrinkled zones requires continuity of radial stress  \cite{Davidovitch11}, we consider 
an axisymmetric deformation of a sheet of radius $L$ on a sphere of radius $R$, subjected to tensile load: $T_L = \sigma_{rr}(L) =  2Y(\frac{W^2}{R^2}) \sqrt{\eK}  (\frac{W}{L} - 1)$. The energy stored in this strained core is \cite{Grason13, Azadi14}: 
\begin{eqnarray}
U_{\rm core} &=& \pi L^2 \frac{T_L^2}{Y} \big(1 + \frac{1}{384} \alpha^2\big)    \nonumber  \\ &\approx&   (4\pi) Y \frac{W^6}{R^4} \eK  \Big[ 1 + \frac{1}{384} \left(\frac{L}{W}\right)^6 \frac{1}{4 \eK}\Big]   \nonumber \ , \\
&\approx& U_{\rm core}^0 +  \pi Y \frac{W^6}{R^4} \frac{1}{384} \Big(\frac{L}{W}\Big)^6 \ , 
\label{eq:ecore}
\end{eqnarray}  
where $\alpha \equiv \tfrac{Y}{T_L}\tfrac{L^2}{R^2}$, and $U_{\rm core}^0$ does not depend on $L$. In the last equation, we ignored the small contribution to the bending cost at the core, and used the approximation, $T_L  \approx 2Y(\frac{W^2}{R^2}) \sqrt{\eK} \cdot \frac{W}{L}$, employing the fact that $\tfrac{W}{L} \gg 1$ in the limit (\ref{eq:inequality}), and recalling that our energy evaluation~(\ref{eq:Ustrain-1}) consists only of leading terms in the expansion.  
\\

Considering the $L$-dependent terms in Eqs.~(\ref{eq:USing},\ref{eq:UReg},\ref{eq:ecore}), we find that the energy is minimized at the value $L$ for which the singular part $U_{\rm sing}$ of the strain energy, and the energy $U_{\rm core}$ of the strained core, are in balance. This yields:  
\begin{equation}
L \to 2^{4/3} \eK^{1/6} W  \ , 
\label{eq:Lfin}
\end{equation}  
and one can check that at this value of $L$: $U_{\rm sing} \sim O(-\eK \log\eK)$, and $U_{\rm core} \sim O(\eK)$, whereas the $L$-dependent contributions to $U_{\rm reg}$ are $O(\eK^{7/6})$ or higher. We thus obtain the final expansion of the strain (with the contribution of the unwrinkled core): 
\begin{multline}
U_{\rm strain} = \pi Y\frac{W^6}{R^4}  \Big\{ \  \frac{2}{3} \ \eK  \big[ - \log\eK + 1 - 8 \log 2 \big]  \\  +  \frac{1}{12} \sqrt{\eoK} \  \ + \cdots \ \Big\} \ .
\label{eq:Ustrain-2}  
\end{multline}

We take a note of the unusual balance underlying the core radius $L$. If the residual strain in the wrinkled zone is totally ignored, a core size $L$ would derive from a  
balance between a core energy $U_{\rm core} \sim Y\tfrac{L^6}{R^4}$ (corresponding to strain $\sim (L/R)^2$ in area $\sim L^2$), whose minimization favors small $L$,   
and the small-$L$ correction to the dominant energy, Eq.~(\ref{eq:limit-energy}) due to bending and substrate deformation, which clearly favors a large $L$. One can see from Eqs.~(\ref{eq:UReg}) that such a balance would give rise to $L \sim \eK^{1/4}$, namely, a smaller core. In contrast, the actual balance underlying Eq.~(\ref{eq:Lfin}) is between two sub-dominant contributions to the energy ($\sim O=\eK$), being ``ignorant'' of the leading energy. This leads to a peculiar situation: a small wrinkle wavelength, $\lambda \sim W \eoK^{1/4}$, is derived from a balance between dominant energies (bending and substrate), whereas a much larger scale, $L \sim W \eK^{1/6} \gg \lambda$, reflects a balance between sub-dominant contributions to the energy.    

The origin of this non-intuitive phenomenology consists of two intimately related features:  
(1) the singular nature of the residual stress ($\sigma_{rr} \sim 1/r$), which corresponds to a singularity in $U_{\rm strain}$ as $L \to 0$; and (2) the suppression of wrinkle amplitude in the vicinity of the core ($f \sim r$, Eq.~\ref{eq:vanish-eps-tt}), which implies a rather small effect of the core size on bending and substrate deformation energies. The combined effect of properties (1) and (2) is an unusual example of {\emph{stress focusing}}, in which the energy density $u_{\rm strain}$ is negligible in comparison to $u_{\rm bend} + u_{\rm sub}$ in the vast part of the wrinkled zone, but becomes large as $r \to 0$, thus governing the core size. 
We expect that such a strain-dominated, stress focusing mechanism that governs the size of an unwrinkled core zone, is relevant to any GIC problem of a thin sheet onto topography with positive Gaussian curvature. 
\subsection{Summary: ITFT for the spherical Winkler model}
Here, we summarize the essential predictions of the ITFT for the asymptotic stress profile in the limit of Eq.~(\ref{eq:inequality}), which we use as a basis for comparison with numerical simulation results in the main text Fig. 2.  We summarize only the leading order terms $O \eK$ in the stress whose derivation was given above,
\begin{equation}
\sigma_{rr}  \to\frac{Y W^2}{R^2} \times \left\{\begin{array}{ll} \frac{3 \eK^{1/3}}{16^{1/3}} - \frac{r^2}{16 W^2} -2 \eK^{1/2}, &  r< L \\  \\ 2 \sqrt{  \eK} \Big( \frac{W}{r} -1 \Big) & \ r\geq L \end{array} \right .
\end{equation}
and  
\begin{equation}
\sigma_{\q\q}  \to \frac{Y W^2}{R^2} \times \left\{\begin{array}{ll}   \frac{3 \eK^{1/3}}{16^{1/3}} - \frac{3r^2}{16 W^2} -2 \eK^{1/2}  &  r< L \\  \\ -2 \sqrt{  \eK}  & \ r\geq L \end{array} \right .
\end{equation}
with $L = 2^{4/3} W \eK^{1/6}$.  We note that,  although continuity of $\sigma_{\q \q} $ is not imposed as condition of the ITFT equations, the minimal energy value of core size leads to continuity of hoop stress at the edge of the wrinkled zone, that is, $\sigma_{\q \q} (L_-) =\sigma_{\q \q} (L_+) =  -2 \frac{Y W^2}{R^2}  \eK^{1/2}$, a feature also found in application of TFT to some other radially symmetric setups~\cite{Davidovitch11}.

\subsection{Accommodating a spatially-constant wavelength}  
Minimization of the dominant contribution to the energy, Eqs.~(\ref{eq:expansion-energy},\ref{eq:limit-energy-o},\ref{eq:limit-energy}), 
implies a constant wrinkle wavelength, $\lambda$, and consequently a radially-varying number of wrinkles, $m(r)=2\pi r/\lambda$, Eq.~(\ref{eq:m-scale-n}), for which the energies $\ubend$ and $\usub$ are in balance. 
Hence, a spatially-constant wrinkle number, 
$m(r)=m_0$ (which has been often assumed in previous studies \cite{Grason13, Hohlfeld15}), is associated with some deviation from this energetically-favorable value. 
Indeed, minimizing $\int_0^{2\pi}d\q\int_0^W rdr  \ (\ubend + u_{\rm sub})$ over wrinkle patterns with 
$m(r)=m_0$, 
leads to an energy larger by $\approx 10\%$ from Eq.~(\ref{eq:limit-energy}). Indeed, we do expect that deviations from a spatially-uniform $m$, 
require an additional energy cost \cite{Paulsen16}, and thereby higher order terms in $\eK,\eoK$, in comparison to the leading terms~(\ref{eq:expansion-energy},\ref{eq:limit-energy}); however, such corrections should not affect the leading-order analysis, which underlies the ITFT equations. This expectation is substantiated by the excellent agreement between the ITFT predictions and the stress profile, energy, and the core size, $L$ observed in simulations (Fig.~2 of the main text). Furthermore, the negligibility of the energetic cost for spatially-varying $m(r)$ is in agreement with a rigorous energetic bound obtained recently by Bella and Kohn \cite{Bella17}, (where the sole expansion parameter -- in the terminology of Eq.~(\ref{eq:inequality}) -- was $\eoK$). 

\section{The spherical stamping problem} 
\subsection{A 3-zone wrinkle pattern}
In contrast to the Winkler model, whereby the substrate stiffness 
$\Ksub$ is explicitly given as a system parameter,  
the spherical stamping problem does not consist of any energetic cost for deforming the spherical topography imposed on the sheet. Instead, the energy consists only of bending and straining the sheet, subjected to the constraint that the deflection of the sheet from the sphere is smaller than the gap, namely: $|\zeta(r,\theta) - \zeta_{sph}(r)| \leq \delta$. 
One can address such a {\emph{non-holonomic}} constraint by assuming a pattern of radial wrinkles, 
$\zeta(r,\theta) \approx \zeta_{sph}(r) + f(r)\cos(m\theta)$, that effectively relaxes the geometrically-induced confinement of latitudes. However, in contrast to the ansatz~(\ref{eq:wrinkle-form}), such a pattern consists of two qualitatively-different annular zones: 

 {\emph{(i)}} In the vicinity of the edge, $L_1<r<W$, where confinement ({\emph{i.e.}} $\tDelta(r)$) is the strongest, the amplitude $f(r)$ exploits the possible deflection by filling the gap, hence $f(r) = \delta$. As was mentioned in the main text, one can think of such a response by imagining a Winkler substrate with spatially-varying stiffness, $\Ksub(r) \sim \tDelta^2(r) (B/\delta^4)$ (which is obtained by requiring Eq. 11 of the main text to yield $\lambda(r)\approx \pi  \delta/\sqrt{\tDelta(r)}$, and consequently yield through the slaving condition ($\pi f/\lambda\approx \sqrt{\tDelta(r)}$) the desired amplitude $f(r) \approx \delta$. (In this sense, $\Ksub(r)$ corresponds to Lagrange multiplier generated by a {\emph{physical boundary}}, namely the infinity rigid spherical walls).

 {\emph{(ii)}} In an inner annulus, $L_2<r<L_1$, where the confinement $\tDelta(r)$ if weaker, but still larger than $\sqrt{B{\Keff}}/Y$ (which is the threshold for hoop buckling), the wrinkles do not fill the gap. Instead, the dominant component in the effective stiffness, $\Keff$ (Eq.~14 of the main text) is the curvature-induced stiffness, $\Kcurv = Y/R^2$ \cite{Paulsen16}, and hence the wavelength of wrinkles in this zone is $\lambda \sim \sqrt{t  R}$, and the amplitude $f(r) \approx \sqrt{\tDelta(r)}\lambda/\pi < \delta$.

\noindent The border, $r\approx L_1$, between the two annular zones, can be found by equating the above expressions for $\Kcurv$ and $\Ksub$, yielding: 
\begin{equation} \tDelta(L_1) \approx \sqrt{\frac{Y}{B}}\frac{\delta^2}{R} \sim  \frac{\delta^2}{tR} \ . 
\label{eq:L1}
\end{equation}
(As was shown in the main text, a straightforward mapping to the spherical Winkler model yields the size, $L_2\sim W(t/\delta)^{2/3}$, of the unwrinkled core).

\subsection{Asymptotic analysis} 
In order to fully characterize the wrinkle pattern, namely -- the exact radii, $L_1,L_2$, of the annular zones, and the exact wavelength, $\lambda(r)$ and amplitude, $f(r)$, we need to find the confinement, $\tDelta(r)$, for a given pair of control parameters, $\epsilon, \chi$ (whose dependence on the physical parameters of the spherical stamping problem is given in Eq.~3 of the main text). A full solution of this problem requires us solve the ITFT equations in each annular zone, and then ``stitch'' the two solutions through appropriate matching conditions (continuity of radial displacement and stress) at $r=L_1$ and $r=L_2$. The resulting equations (which fully characterize the radial confinement $\tDelta(r)$ as well as the radii, $L_1,L_2$) are cumbersome, but rather straightforward, and will be described elsewhere; their solution underlies the energy and force plots (Fig.~4 of the main text). However, one can obtain the asymptotic solution in the limit~(\ref{eq:inequality}) by substituting in Eq.~(\ref{eq:L1}) the asymptotic   
confinement function, $\tDelta(r) \to \tfrac{1}{6}(r/R)^2$, which is required for asymptotic elimination of radial strain. We thus obtain that:
\begin{equation} L_1 \approx \sqrt{6} \delta \sqrt{R/t} \  . 
\label{eq:L1s}
\end{equation}
Obviously, in order for a ``2-zone'' wrinkle pattern to be a valid solution of the ITFT equations, $L_1$ must satisfy the double inequality: 
\begin{equation}  
L_2 <L_1 <W 
\label{eq:L1-ineq}
\end{equation}
If the upper boundary is approached ($L_1 \to W$), the wrinkles no longer fill the gap (namely, the whole pattern is limited by the curvature of the shell rather than by the gap). Using Eq.~(\ref{eq:L1s}) we find that this occurs when $\delta \to \delta_c$ where, $\delta_c \sim W\sqrt{t/R}$, as reported in the main text. For $\delta>\delta_c$, the stamping shells do not exert any force on the wrinkled sheet and the stamping force $F(\delta)$ vanishes (end of curves in Fig.~4 of the main text).  Beyond this point we expect that another ({\emph{i.e.}} non-wrinkly) type of energy minimizer emerges (purple regions in the phase diagram, Fig.~4B of the main text). 

If the lower boundary in Eq.~(\ref{eq:L1-ineq}) is approached ($L_1 \to L_2 $), the 2-zone wrinkle pattern becomes a ``standard'' wrinkle pattern, where wrinkles fill the gap except at an unwrinkled core (of radius $\approx L_2 \sim W (t/\delta)^{2/3}$), similarly to the spherical Winkler problem (albeit with $\Ksub \to (\tfrac{1}{6})^2 (r/R)^4 B/\delta^4$). Inspection of Eq.~(\ref{eq:L1}) shows that this happens for $\delta<\bar{\delta}$, where $\bar{\delta} \sim W^{3/5} R^{-3/10} t^{7/10}$. The corresponding parameter regime is thus, $ O(t) <\delta < \bar{\delta}$ (red zone in the phase diagram, Fig.~4B of the main text). 

\begin{figure}
\centering
\includegraphics[width=0.5\textwidth]{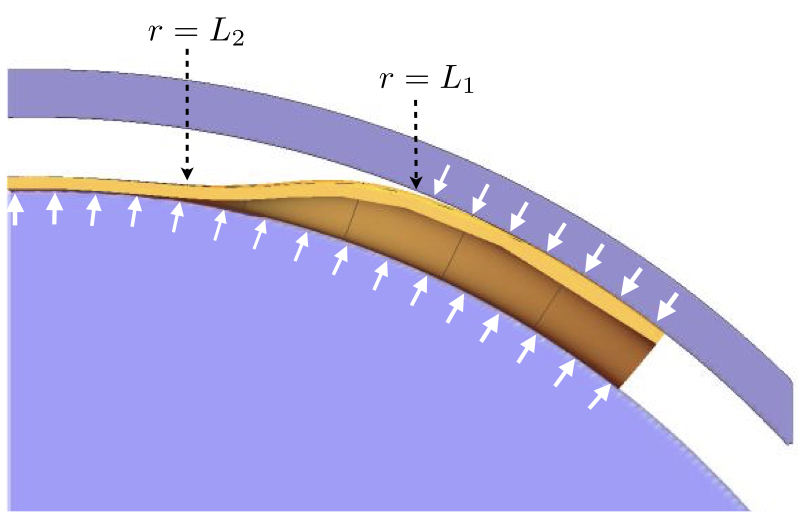}
\caption{\label{fig:1}  A schematic of a radial slice of the sheet (orange) in the spherical stamp (purple), in a 3-zone radial wrinkled state. The sheet remains in contact with the bottom stamp along its radial profile, resulting in a upward contact force, while the sheet only makes contact with the top stamp in the gap filling region, $r > L_1$, which contributes a balancing downward force.  The white arrows illustrate azimuthally-averaged force from the top and bottom stamps, with the thickness of the arrows a schematic illustration of the variation of magnitudes ({\emph{e.g.}} the net downward force is largest as $r \to W$). }
\end{figure}

\subsection{Confining forces} To understand the nature of the confining forces generated by the stamp, first consider the limit $\delta \to t$, where the stamp restricts the sheet to an axisymmetric, perfectly spherical shape (i.e. $\zeta \to - r^2/(2R)$). The pressure generated by stamp must balance the membranal resistance of the sheet: $(\srr(r) + \sqq(r))/R$.  For this axisymmetric limit, the stress is known analytically~\cite{Azadi12}, and satisfies: $\srr(r) + \sqq(r)>0$ (tensile) for $r <  W/\sqrt{2}$ implying a net upward force from the stamp in this inner region, while  $\srr(r) + \sqq(r)<0$ (compressive) for  $r > W/\sqrt{2} $ which must be balanced by a net downward force from the stamp.

Fig.~\ref{fig:1} shows a schematic of qualitatively similar pattern of confining forces of the stamp acting on a 3-zone wrinkled pattern ($t\ll \delta \ll W^2/R$).  In the tensile core ($r<L_2$) the sheet is in contact with the bottom shell, and the corresponding normal force $f_{bot}(r)$ (per area) exerted by the shell is balanced by a downward membranal force generated by tension, $(\srr(r) + \sqq(r))/R>0$. In contrast, in the wrinkled zone the normal forces exerted by the confining shells are localized at contact lines with the sheet -- valleys or crests of the wrinkles. In order to understand the nature of these forces, let us consider first the gap-filling region, $L_1<r<W$. Here, the oppositely-oriented normal forces exerted by the bottom and top shells, $f_{bot}(r)$ and $f_{top}(r)$, respectively, generate an azimuthally-oscillating force pattern that -- in tandem with the bending force, $ \sim B (m/r)^4 \delta \cdot \cos(m\theta)$ -- balance the (destabilizing) compressive force, $- \sqq  (m/r)^2 \delta  \cdot \cos(m\theta)$. The analogy between the gap-filling wrinkled zone and a spherical Winkler foundation with spring constant $\Ksub \sim B/\delta^4$ implies that the magnitude of $f_{bot},f_{top} \sim B/\delta^3$. Additionally, the radial tension and curvature in the sheet give rise to a net ({\emph{i.e.}} non-oscillatory) force, $(\srr + \sqq) /R\big|_{r \lesssim W} \sim - B/(\delta^2 R)$, which acts to flatten the sheet. To balance this force, the downward force exerted by the top shell force must be slightly larger than the upward force from the bottom: $(f_{top}- f_{bot})/f_{top} \sim \delta/R \ll 1$. 

Let us turn now to the zone $L_2<r<L_1$. Here, wrinkles do not fill the gap, making contact only with the bottom shell, namely $f_{bot} >0, f_{top}=0$, such that the confining forces do not generate an azimuthally-oscillatory pattern but only a net (upward) force $f_{bot}$. However, in this zone the effective stiffness $\Keff \approx \Kcurv$ that stabilizes an oscillatory wrinkle pattern and determines the wavelength, $\lambda \sim \sqrt{Rt}$, does not stem from external forces, but rather from the oscillatory part of the radial stress, $\srr$ (see discussion following Eq.~(\ref{eq:stress-asy-2}) and Eq. 14 of the main text). Hence, while $f_{top}=0$, the force $f_{bot}$ is again determined through balance with the non-oscillatory part of the  stress, $f_{bot} \sim (\srr+ \sqq) /R >0$.

\bibliography{Gauss-Euler}  

\begin{thebibliography}{28}%
\makeatletter
\providecommand \@ifxundefined [1]{%
 \@ifx{#1\undefined}
}%
\providecommand \@ifnum [1]{%
 \ifnum #1\expandafter \@firstoftwo
 \else \expandafter \@secondoftwo
 \fi
}%
\providecommand \@ifx [1]{%
 \ifx #1\expandafter \@firstoftwo
 \else \expandafter \@secondoftwo
 \fi
}%
\providecommand \natexlab [1]{#1}%
\providecommand \enquote  [1]{``#1''}%
\providecommand \bibnamefont  [1]{#1}%
\providecommand \bibfnamefont [1]{#1}%
\providecommand \citenamefont [1]{#1}%
\providecommand \href@noop [0]{\@secondoftwo}%
\providecommand \href [0]{\begingroup \@sanitize@url \@href}%
\providecommand \@href[1]{\@@startlink{#1}\@@href}%
\providecommand \@@href[1]{\endgroup#1\@@endlink}%
\providecommand \@sanitize@url [0]{\catcode `\\12\catcode `\$12\catcode
  `\&12\catcode `\#12\catcode `\^12\catcode `\_12\catcode `\%12\relax}%
\providecommand \@@startlink[1]{}%
\providecommand \@@endlink[0]{}%
\providecommand \url  [0]{\begingroup\@sanitize@url \@url }%
\providecommand \@url [1]{\endgroup\@href {#1}{\urlprefix }}%
\providecommand \urlprefix  [0]{URL }%
\providecommand \Eprint [0]{\href }%
\providecommand \doibase [0]{http://dx.doi.org/}%
\providecommand \selectlanguage [0]{\@gobble}%
\providecommand \bibinfo  [0]{\@secondoftwo}%
\providecommand \bibfield  [0]{\@secondoftwo}%
\providecommand \translation [1]{[#1]}%
\providecommand \BibitemOpen [0]{}%
\providecommand \bibitemStop [0]{}%
\providecommand \bibitemNoStop [0]{.\EOS\space}%
\providecommand \EOS [0]{\spacefactor3000\relax}%
\providecommand \BibitemShut  [1]{\csname bibitem#1\endcsname}%
\let\auto@bib@innerbib\@empty
\bibitem [{\citenamefont {Levien}(2008)}]{ElasticaReview08}%
  \BibitemOpen
  \bibfield  {author} {\bibinfo {author} {\bibfnamefont {R.}~\bibnamefont
  {Levien}},\ }\href@noop {} {\emph {\bibinfo {title} {The elastica: a
  mathematical history}}},\ \bibinfo {type} {Tech. Rep.}\ (\bibinfo
  {institution} {University of California Berkeley},\ \bibinfo {year}
  {2008})\BibitemShut {NoStop}%
\bibitem [{\citenamefont {Landau}\ and\ \citenamefont
  {Lifschitz}(1986)}]{LL86}%
  \BibitemOpen
  \bibfield  {author} {\bibinfo {author} {\bibfnamefont {L.~D.}\ \bibnamefont
  {Landau}}\ and\ \bibinfo {author} {\bibfnamefont {E.~M.}\ \bibnamefont
  {Lifschitz}},\ }\href@noop {} {\emph {\bibinfo {title} {The Theory of
  Elasticity}}}\ (\bibinfo  {publisher} {Pergamon},\ \bibinfo {year}
  {1986})\BibitemShut {NoStop}%
\bibitem [{\citenamefont {Mansfield}(1989)}]{Mansfield}%
  \BibitemOpen
  \bibfield  {author} {\bibinfo {author} {\bibfnamefont {E.~H.}\ \bibnamefont
  {Mansfield}},\ }\href@noop {} {\emph {\bibinfo {title} {The Bending and
  Stretching of Plates}}}\ (\bibinfo  {publisher} {Cambridge University
  Press},\ \bibinfo {year} {1989})\BibitemShut {NoStop}%
\bibitem [{\citenamefont {Cerda}\ and\ \citenamefont
  {Mahadevan}(1998)}]{Cerda98}%
  \BibitemOpen
  \bibfield  {author} {\bibinfo {author} {\bibfnamefont {E.}~\bibnamefont
  {Cerda}}\ and\ \bibinfo {author} {\bibfnamefont {L.}~\bibnamefont
  {Mahadevan}},\ }\href@noop {} {\bibfield  {journal} {\bibinfo  {journal}
  {Phys. Rev. Lett.}\ }\textbf {\bibinfo {volume} {80}},\ \bibinfo {pages}
  {2358} (\bibinfo {year} {1998})}\BibitemShut {NoStop}%
\bibitem [{\citenamefont {Witten}(2007)}]{Witten07}%
  \BibitemOpen
  \bibfield  {author} {\bibinfo {author} {\bibfnamefont {T.~A.}\ \bibnamefont
  {Witten}},\ }\href@noop {} {\bibfield  {journal} {\bibinfo  {journal} {Rev.
  Mod. Phys.}\ }\textbf {\bibinfo {volume} {79}},\ \bibinfo {pages} {643}
  (\bibinfo {year} {2007})}\BibitemShut {NoStop}%
\bibitem [{\citenamefont {Hure}\ \emph {et~al.}(2012)\citenamefont {Hure},
  \citenamefont {Roman},\ and\ \citenamefont {Bico}}]{Hure12}%
  \BibitemOpen
  \bibfield  {author} {\bibinfo {author} {\bibfnamefont {J.}~\bibnamefont
  {Hure}}, \bibinfo {author} {\bibfnamefont {B.}~\bibnamefont {Roman}}, \ and\
  \bibinfo {author} {\bibfnamefont {J.}~\bibnamefont {Bico}},\ }\href@noop {}
  {\bibfield  {journal} {\bibinfo  {journal} {Phys. Rev. Lett.}\ }\textbf
  {\bibinfo {volume} {109}},\ \bibinfo {pages} {054302} (\bibinfo {year}
  {2012})}\BibitemShut {NoStop}%
\bibitem [{\citenamefont {Aharoni}\ \emph {et~al.}(2017)\citenamefont
  {Aharoni}, \citenamefont {Todorova}, \citenamefont {O}, \citenamefont
  {Goehring}, \citenamefont {Kamien},\ and\ \citenamefont
  {Katifori}}]{Aharoni17}%
  \BibitemOpen
  \bibfield  {author} {\bibinfo {author} {\bibfnamefont {H.}~\bibnamefont
  {Aharoni}}, \bibinfo {author} {\bibfnamefont {D.~V.}\ \bibnamefont
  {Todorova}}, \bibinfo {author} {\bibfnamefont {A.}~\bibnamefont {O}},
  \bibinfo {author} {\bibfnamefont {L.}~\bibnamefont {Goehring}}, \bibinfo
  {author} {\bibfnamefont {R.~D.}\ \bibnamefont {Kamien}}, \ and\ \bibinfo
  {author} {\bibfnamefont {E.}~\bibnamefont {Katifori}},\ }\href@noop {}
  {\bibfield  {journal} {\bibinfo  {journal} {Nature Comm.}\ }\textbf {\bibinfo
  {volume} {8}},\ \bibinfo {pages} {15809} (\bibinfo {year}
  {2017})}\BibitemShut {NoStop}%
\bibitem [{\citenamefont {Grason}\ and\ \citenamefont
  {Davidovitch}(2013)}]{Grason13}%
  \BibitemOpen
  \bibfield  {author} {\bibinfo {author} {\bibfnamefont {G.~M.}\ \bibnamefont
  {Grason}}\ and\ \bibinfo {author} {\bibfnamefont {B.}~\bibnamefont
  {Davidovitch}},\ }\href@noop {} {\bibfield  {journal} {\bibinfo  {journal}
  {Proc. Natl. Acad. Sci.}\ }\textbf {\bibinfo {volume} {110}},\ \bibinfo
  {pages} {12893} (\bibinfo {year} {2013})}\BibitemShut {NoStop}%
\bibitem [{\citenamefont {Hohlfeld}\ and\ \citenamefont
  {Davidovitch}(2015)}]{Hohlfeld15}%
  \BibitemOpen
  \bibfield  {author} {\bibinfo {author} {\bibfnamefont {E.}~\bibnamefont
  {Hohlfeld}}\ and\ \bibinfo {author} {\bibfnamefont {B.}~\bibnamefont
  {Davidovitch}},\ }\href@noop {} {\bibfield  {journal} {\bibinfo  {journal}
  {Phys. Rev. E}\ }\textbf {\bibinfo {volume} {91}},\ \bibinfo {pages} {012407}
  (\bibinfo {year} {2015})}\BibitemShut {NoStop}%
\bibitem [{\citenamefont {Bella}\ and\ \citenamefont {Kohn}(2017)}]{Bella17}%
  \BibitemOpen
  \bibfield  {author} {\bibinfo {author} {\bibfnamefont {P.}~\bibnamefont
  {Bella}}\ and\ \bibinfo {author} {\bibfnamefont {R.~V.}\ \bibnamefont
  {Kohn}},\ }\href@noop {} {\bibfield  {journal} {\bibinfo  {journal} {Phil.
  Trans. R. Soc. A}\ }\textbf {\bibinfo {volume} {375}},\ \bibinfo {pages}
  {20160157} (\bibinfo {year} {2017})}\BibitemShut {NoStop}%
\bibitem [{\citenamefont {Seung}\ and\ \citenamefont
  {Nelson}(1988)}]{NelsonSeung}%
  \BibitemOpen
  \bibfield  {author} {\bibinfo {author} {\bibfnamefont {H.~S.}\ \bibnamefont
  {Seung}}\ and\ \bibinfo {author} {\bibfnamefont {D.~R.}\ \bibnamefont
  {Nelson}},\ }\href@noop {} {\bibfield  {journal} {\bibinfo  {journal} {Phys.
  Rev. A}\ }\textbf {\bibinfo {volume} {38}},\ \bibinfo {pages} {1005}
  (\bibinfo {year} {1988})}\BibitemShut {NoStop}%
\bibitem [{\citenamefont {Vernizzi}\ \emph {et~al.}(2011)\citenamefont
  {Vernizzi}, \citenamefont {Sknepnek},\ and\ \citenamefont {de~la
  Cruz}}]{Rastko}%
  \BibitemOpen
  \bibfield  {author} {\bibinfo {author} {\bibfnamefont {G.}~\bibnamefont
  {Vernizzi}}, \bibinfo {author} {\bibfnamefont {R.}~\bibnamefont {Sknepnek}},
  \ and\ \bibinfo {author} {\bibfnamefont {M.~O.}\ \bibnamefont {de~la Cruz}},\
  }\href@noop {} {\bibfield  {journal} {\bibinfo  {journal} {Proc. Nat. Aca.
  Sci. USA}\ }\textbf {\bibinfo {volume} {108}},\ \bibinfo {pages} {4292}
  (\bibinfo {year} {2011})}\BibitemShut {NoStop}%
\bibitem [{\citenamefont {Bowden}\ \emph {et~al.}(1998)\citenamefont {Bowden},
  \citenamefont {Brittain}, \citenamefont {Evans}, \citenamefont {Hutchinson},\
  and\ \citenamefont {Whiteside}}]{Bowden98}%
  \BibitemOpen
  \bibfield  {author} {\bibinfo {author} {\bibfnamefont {N.}~\bibnamefont
  {Bowden}}, \bibinfo {author} {\bibfnamefont {S.}~\bibnamefont {Brittain}},
  \bibinfo {author} {\bibfnamefont {A.~G.}\ \bibnamefont {Evans}}, \bibinfo
  {author} {\bibfnamefont {J.~W.}\ \bibnamefont {Hutchinson}}, \ and\ \bibinfo
  {author} {\bibfnamefont {G.~M.}\ \bibnamefont {Whiteside}},\ }\href@noop {}
  {\bibfield  {journal} {\bibinfo  {journal} {Nature}\ }\textbf {\bibinfo
  {volume} {393}},\ \bibinfo {pages} {146} (\bibinfo {year}
  {1998})}\BibitemShut {NoStop}%
\bibitem [{\citenamefont {Cerda}\ and\ \citenamefont
  {Mahadevan}(2003)}]{Cerda03}%
  \BibitemOpen
  \bibfield  {author} {\bibinfo {author} {\bibfnamefont {E.}~\bibnamefont
  {Cerda}}\ and\ \bibinfo {author} {\bibfnamefont {L.}~\bibnamefont
  {Mahadevan}},\ }\href@noop {} {\bibfield  {journal} {\bibinfo  {journal}
  {Phys. Rev. Lett.}\ }\textbf {\bibinfo {volume} {90}},\ \bibinfo {pages}
  {074302} (\bibinfo {year} {2003})}\BibitemShut {NoStop}%
\bibitem [{\citenamefont {Davidovitch}\ \emph {et~al.}(2011)\citenamefont
  {Davidovitch}, \citenamefont {Schroll}, \citenamefont {Vella}, \citenamefont
  {Adda-Bedia},\ and\ \citenamefont {Cerda}}]{Davidovitch11}%
  \BibitemOpen
  \bibfield  {author} {\bibinfo {author} {\bibfnamefont {B.}~\bibnamefont
  {Davidovitch}}, \bibinfo {author} {\bibfnamefont {R.~D.}\ \bibnamefont
  {Schroll}}, \bibinfo {author} {\bibfnamefont {D.}~\bibnamefont {Vella}},
  \bibinfo {author} {\bibfnamefont {M.}~\bibnamefont {Adda-Bedia}}, \ and\
  \bibinfo {author} {\bibfnamefont {E.}~\bibnamefont {Cerda}},\ }\href@noop {}
  {\bibfield  {journal} {\bibinfo  {journal} {Proc. Natl. Acad. Sci. USA}\
  }\textbf {\bibinfo {volume} {108}},\ \bibinfo {pages} {18227} (\bibinfo
  {year} {2011})}\BibitemShut {NoStop}%
\bibitem [{\citenamefont {Paulsen}\ \emph {et~al.}(2016)\citenamefont
  {Paulsen}, \citenamefont {Hohlfeld}, \citenamefont {King}, \citenamefont
  {Huang}, \citenamefont {Qiu}, \citenamefont {Russell}, \citenamefont {Menon},
  \citenamefont {Vella},\ and\ \citenamefont {Davidovitch}}]{Paulsen16}%
  \BibitemOpen
  \bibfield  {author} {\bibinfo {author} {\bibfnamefont {J.~D.}\ \bibnamefont
  {Paulsen}}, \bibinfo {author} {\bibfnamefont {E.}~\bibnamefont {Hohlfeld}},
  \bibinfo {author} {\bibfnamefont {H.}~\bibnamefont {King}}, \bibinfo {author}
  {\bibfnamefont {J.~S.}\ \bibnamefont {Huang}}, \bibinfo {author}
  {\bibfnamefont {Z.}~\bibnamefont {Qiu}}, \bibinfo {author} {\bibfnamefont
  {T.~P.~R.}\ \bibnamefont {Russell}}, \bibinfo {author} {\bibfnamefont
  {N.}~\bibnamefont {Menon}}, \bibinfo {author} {\bibfnamefont
  {D.}~\bibnamefont {Vella}}, \ and\ \bibinfo {author} {\bibfnamefont
  {B.}~\bibnamefont {Davidovitch}},\ }\href@noop {} {\bibfield  {journal}
  {\bibinfo  {journal} {Proc. Nat. Aca. Sci. USA}\ }\textbf {\bibinfo {volume}
  {113}},\ \bibinfo {pages} {1144} (\bibinfo {year} {2016})}\BibitemShut
  {NoStop}%
\bibitem [{\citenamefont {Audoly}\ and\ \citenamefont
  {Boudaoud}(2008)}]{Audoly08}%
  \BibitemOpen
  \bibfield  {author} {\bibinfo {author} {\bibfnamefont {B.}~\bibnamefont
  {Audoly}}\ and\ \bibinfo {author} {\bibfnamefont {A.}~\bibnamefont
  {Boudaoud}},\ }\href@noop {} {\bibfield  {journal} {\bibinfo  {journal} {J.
  Mech. Phy. Solids}\ }\textbf {\bibinfo {volume} {56}},\ \bibinfo {pages}
  {2401} (\bibinfo {year} {2008})}\BibitemShut {NoStop}%
\bibitem [{\citenamefont {Kohn}\ and\ \citenamefont {Nguyen}(2013)}]{Nguyen13}%
  \BibitemOpen
  \bibfield  {author} {\bibinfo {author} {\bibfnamefont {R.~V.}\ \bibnamefont
  {Kohn}}\ and\ \bibinfo {author} {\bibfnamefont {H.~M.}\ \bibnamefont
  {Nguyen}},\ }\href@noop {} {\bibfield  {journal} {\bibinfo  {journal} {J.
  Nonlin. Sci.}\ }\textbf {\bibinfo {volume} {23}},\ \bibinfo {pages} {343}
  (\bibinfo {year} {2013})}\BibitemShut {NoStop}%
\bibitem [{\citenamefont {King}\ \emph {et~al.}(2012)\citenamefont {King},
  \citenamefont {Schroll}, \citenamefont {Davidovitch},\ and\ \citenamefont
  {Menon}}]{King12}%
  \BibitemOpen
  \bibfield  {author} {\bibinfo {author} {\bibfnamefont {H.}~\bibnamefont
  {King}}, \bibinfo {author} {\bibfnamefont {R.~D.}\ \bibnamefont {Schroll}},
  \bibinfo {author} {\bibfnamefont {B.}~\bibnamefont {Davidovitch}}, \ and\
  \bibinfo {author} {\bibfnamefont {N.}~\bibnamefont {Menon}},\ }\href@noop {}
  {\bibfield  {journal} {\bibinfo  {journal} {Proc. Natl. Acad. Sci. USA}\
  }\textbf {\bibinfo {volume} {109}},\ \bibinfo {pages} {9716} (\bibinfo {year}
  {2012})}\BibitemShut {NoStop}%
\bibitem [{\citenamefont {Pocivavsek}\ \emph {et~al.}(2008)\citenamefont
  {Pocivavsek}, \citenamefont {Dellsy}, \citenamefont {Kern}, \citenamefont
  {Johnson}, \citenamefont {Lin}, \citenamefont {Lee},\ and\ \citenamefont
  {Cerda}}]{Pocivavsek08}%
  \BibitemOpen
  \bibfield  {author} {\bibinfo {author} {\bibfnamefont {L.}~\bibnamefont
  {Pocivavsek}}, \bibinfo {author} {\bibfnamefont {R.}~\bibnamefont {Dellsy}},
  \bibinfo {author} {\bibfnamefont {A.}~\bibnamefont {Kern}}, \bibinfo {author}
  {\bibfnamefont {S.}~\bibnamefont {Johnson}}, \bibinfo {author} {\bibfnamefont
  {B.~H.}\ \bibnamefont {Lin}}, \bibinfo {author} {\bibfnamefont {K.~Y.~C.}\
  \bibnamefont {Lee}}, \ and\ \bibinfo {author} {\bibfnamefont
  {E.}~\bibnamefont {Cerda}},\ }\href@noop {} {\bibfield  {journal} {\bibinfo
  {journal} {Science}\ }\textbf {\bibinfo {volume} {320}},\ \bibinfo {pages}
  {912} (\bibinfo {year} {2008})}\BibitemShut {NoStop}%
\bibitem [{\citenamefont {Diamant}\ and\ \citenamefont
  {Witten}(2011)}]{Diamant11}%
  \BibitemOpen
  \bibfield  {author} {\bibinfo {author} {\bibfnamefont {H.}~\bibnamefont
  {Diamant}}\ and\ \bibinfo {author} {\bibfnamefont {T.~A.}\ \bibnamefont
  {Witten}},\ }\href@noop {} {\bibfield  {journal} {\bibinfo  {journal} {Phys.
  Rev. Lett.}\ }\textbf {\bibinfo {volume} {107}},\ \bibinfo {pages} {164302}
  (\bibinfo {year} {2011})}\BibitemShut {NoStop}%
\bibitem [{\citenamefont {Paulsen}\ \emph {et~al.}(2017)\citenamefont
  {Paulsen}, \citenamefont {D\'emery}, \citenamefont {Toga}, \citenamefont
  {Russell}, \citenamefont {Davidovitch},\ and\ \citenamefont
  {Menon}}]{Paulsen17a}%
  \BibitemOpen
  \bibfield  {author} {\bibinfo {author} {\bibfnamefont {J.~D.}\ \bibnamefont
  {Paulsen}}, \bibinfo {author} {\bibfnamefont {V.}~\bibnamefont {D\'emery}},
  \bibinfo {author} {\bibfnamefont {K.~B.}\ \bibnamefont {Toga}}, \bibinfo
  {author} {\bibfnamefont {T.~P.~R.}\ \bibnamefont {Russell}}, \bibinfo
  {author} {\bibfnamefont {B.}~\bibnamefont {Davidovitch}}, \ and\ \bibinfo
  {author} {\bibfnamefont {N.}~\bibnamefont {Menon}},\ }\href@noop {}
  {\bibfield  {journal} {\bibinfo  {journal} {Phys. Rev. Lett.}\ }\textbf
  {\bibinfo {volume} {118}},\ \bibinfo {pages} {048004} (\bibinfo {year}
  {2017})}\BibitemShut {NoStop}%
\bibitem [{\citenamefont {Aharoni}\ and\ \citenamefont
  {Sharon}(2010)}]{Aharoni10}%
  \BibitemOpen
  \bibfield  {author} {\bibinfo {author} {\bibfnamefont {H.}~\bibnamefont
  {Aharoni}}\ and\ \bibinfo {author} {\bibfnamefont {E.}~\bibnamefont
  {Sharon}},\ }\href@noop {} {\bibfield  {journal} {\bibinfo  {journal} {Nat
  Mat}\ }\textbf {\bibinfo {volume} {109}},\ \bibinfo {pages} {054302}
  (\bibinfo {year} {2010})}\BibitemShut {NoStop}%
\bibitem [{\citenamefont {Nash}(1954)}]{Nash}%
  \BibitemOpen
  \bibfield  {author} {\bibinfo {author} {\bibfnamefont {J.}~\bibnamefont
  {Nash}},\ }\href@noop {} {\bibfield  {journal} {\bibinfo  {journal} {Annals
  of Mathematics}\ }\textbf {\bibinfo {volume} {9}},\ \bibinfo {pages}
  {383?396} (\bibinfo {year} {1954})}\BibitemShut {NoStop}%
\bibitem [{\citenamefont {Gemmer}\ \emph {et~al.}(2016)\citenamefont {Gemmer},
  \citenamefont {Sharon}, \citenamefont {Shearman},\ and\ \citenamefont
  {Venkataramani}}]{Gemmer16}%
  \BibitemOpen
  \bibfield  {author} {\bibinfo {author} {\bibfnamefont {J.}~\bibnamefont
  {Gemmer}}, \bibinfo {author} {\bibfnamefont {E.}~\bibnamefont {Sharon}},
  \bibinfo {author} {\bibfnamefont {T.}~\bibnamefont {Shearman}}, \ and\
  \bibinfo {author} {\bibfnamefont {S.~C.}\ \bibnamefont {Venkataramani}},\
  }\href@noop {} {\bibfield  {journal} {\bibinfo  {journal} {Euro. Phys.
  Lett.}\ }\textbf {\bibinfo {volume} {114}},\ \bibinfo {pages} {24003}
  (\bibinfo {year} {2016})}\BibitemShut {NoStop}%
\bibitem [{\citenamefont {Timoshenko}\ and\ \citenamefont
  {Goodier}(1970)}]{timoshenko70}%
  \BibitemOpen
  \bibfield  {author} {\bibinfo {author} {\bibfnamefont {S.~P.}\ \bibnamefont
  {Timoshenko}}\ and\ \bibinfo {author} {\bibfnamefont {J.~N.}\ \bibnamefont
  {Goodier}},\ }\href@noop {} {\emph {\bibinfo {title} {Theory of
  Elasticity}}}\ (\bibinfo  {publisher} {McGraw Hill},\ \bibinfo {year}
  {1970})\BibitemShut {NoStop}%
\bibitem [{\citenamefont {Azadi}\ and\ \citenamefont
  {Grason}(2014{\natexlab{a}})}]{Azadi12}%
  \BibitemOpen
  \bibfield  {author} {\bibinfo {author} {\bibfnamefont {A.}~\bibnamefont
  {Azadi}}\ and\ \bibinfo {author} {\bibfnamefont {G.~M.}\ \bibnamefont
  {Grason}},\ }\href@noop {} {\bibfield  {journal} {\bibinfo  {journal} {Phys.
  Rev. Lett.}\ }\textbf {\bibinfo {volume} {112}},\ \bibinfo {pages} {225502}
  (\bibinfo {year} {2014}{\natexlab{a}})}\BibitemShut {NoStop}%
\bibitem [{\citenamefont {Azadi}\ and\ \citenamefont
  {Grason}(2014{\natexlab{b}})}]{Azadi14}%
  \BibitemOpen
  \bibfield  {author} {\bibinfo {author} {\bibfnamefont {A.}~\bibnamefont
  {Azadi}}\ and\ \bibinfo {author} {\bibfnamefont {G.~M.}\ \bibnamefont
  {Grason}},\ }\href@noop {} {\bibfield  {journal} {\bibinfo  {journal} {Phys.
  Rev. E}\ }\textbf {\bibinfo {volume} {94}},\ \bibinfo {pages} {013003}
  (\bibinfo {year} {2014}{\natexlab{b}})}\BibitemShut {NoStop}%
\end{thebibliography}%

\end{document}